\documentclass{article}
\DeclareUnicodeCharacter{2212}{-}

\usepackage{arxiv}

\usepackage[utf8]{inputenc} 
\usepackage[T1]{fontenc}    
\usepackage{url}
\usepackage{booktabs}       
\usepackage{amsfonts}       
\usepackage{nicefrac}       
\usepackage{microtype}      
\usepackage{lipsum}		
\usepackage{graphicx}
\usepackage{natbib}
\usepackage{doi}
\usepackage[utf8]{inputenc}
\usepackage[T1]{fontenc}
\usepackage{bm}
\usepackage{amsmath}
\usepackage{amsthm}
\usepackage{amssymb}
\usepackage{mathrsfs} 
\usepackage{xcolor}
\usepackage{color}
\usepackage{titlesec}
\usepackage{fancyhdr}
\usepackage{array} 

\usepackage{pifont}
\usepackage{ulem}
\usepackage{amsmath}
\usepackage{multirow}
\usepackage{rotating}
\usepackage{pdflscape}
\newtheoremstyle{definition-style}
  {3pt}
  {3pt}
  {}
  {}
  {\bfseries}
  {:}
  {}
  {\thmname{#1}~\thmnumber{#2}~\thmnote{(#3)}}

\title{Critical Nodes Identification in Complex Networks: A Survey}

\author{{Duxin Chen}  \thanks{Co-author: Contributed equally to this article} \\
	School of Mathematics\\
	Southeast University\\
	Nanjing {\rm211189}, China \\
	\texttt{chendx@seu.edu.cn} \\  
         \and
        {Jiawen Chen}~$^{*}$
        \\
	School of Mathematics\\
	Southeast University\\
	Nanjing {\rm211189}, China \\
	\texttt{jiawenchen@seu.edu.cn} \\
	\and
	{Xiaoyu Zhang} \\
	School of Mathematics\\
	Southeast University\\
	Nanjing {\rm211189}, China \\
	\texttt{220242060@seu.edu.cn} \\
        \and
	{Qinghan Jia} \\
	School of Mathematics\\
	Southeast University\\
	Nanjing {\rm211189}, China \\
		\texttt{220242053@seu.edu.cn} \\ 
        \and
        {Xiaolu Liu} \\
	School of Automation\\
	Nanjing Institute of Technology\\
	Nanjing {\rm211167}, China \\
		\texttt{emspc@126.com} \\ 
        \and
        {Ye Sun } \\
	School of Mathematics\\
	Southeast University\\
	Nanjing {\rm211189}, China \\
		\texttt{yesun@seu.edu.cn} \\ 
        \and
        {Linyuan Lü} \\
	School of Cyber Science and Technology,\\
	University of Science and Technology of China\\
	Hefei {\rm230026}, China \\
		\texttt{linyuan.lv@ustc.edu.cn} \\ 
         \and
        {Wenwu Yu}~\thanks{Corresponding author: wwyu@seu.edu.cn}  \\
	School of Mathematics\\
	Southeast University\\
	Nanjing {\rm211189}, China \\
	\texttt{wwyu@seu.edu.cn} \\      
}

\renewcommand{\shorttitle}{\textit{arXiv} Template}

\begin{document}
\maketitle

\begin{abstract}
Complex networks have become essential tools for understanding diverse phenomena in social systems, traffic systems, biomolecular systems, and financial systems. Identifying critical nodes is a central theme in contemporary research, serving as a vital bridge between theoretical foundations and practical applications. Nevertheless, the intrinsic complexity and structural heterogeneity characterizing real-world networks, with particular emphasis on dynamic and higher-order networks, present substantial obstacles to the development of universal frameworks for critical node identification. This paper provides a comprehensive review of critical node identification techniques, categorizing them into seven main classes: centrality, critical nodes deletion problem, influence maximization, network control, artificial intelligence, higher-order and dynamic methods. Our review bridges the gaps in existing surveys by systematically classifying methods based on their methodological foundations and practical implications, and by highlighting their strengths, limitations, and applicability across different network types. 
Our work enhances the understanding of critical node research by identifying key challenges, such as algorithmic universality, real-time evaluation in dynamic networks, analysis of higher-order structures, and computational efficiency in large-scale networks. The structured synthesis consolidates current progress and highlights open questions, particularly in modeling temporal dynamics, advancing efficient algorithms, integrating machine learning approaches, and developing scalable and interpretable metrics for complex systems.


\end{abstract}

\begin{figure}
    \centering
    \includegraphics[width=1\linewidth]{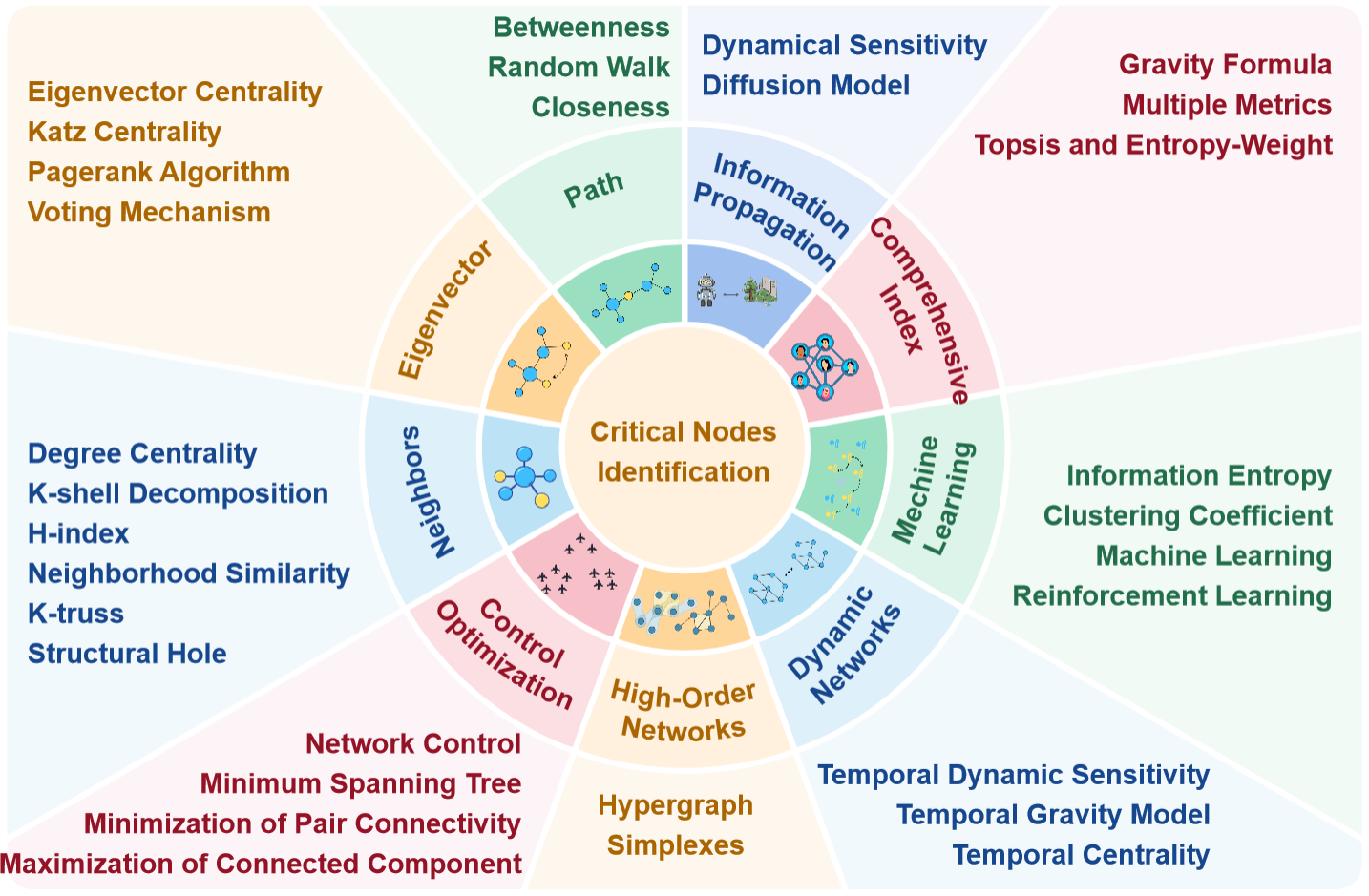}
    \caption{Graphic Abstract.}
    \label{fig001}
\end{figure}

\section{Introduction}
Complex networks and systems are fundamental to diverse fields, including social network analysis, urban traffic modeling~\cite{communicability2018}, biomolecular structure analysis~\cite{de2023more}, and financial system~\cite{bardoscia2021physics}, where understanding the interplay between network structure, system architecture, and functionality is a central research focus~\cite{Yu2024}. A typical network consists of nodes, representing entities within a real system, and edges, which encode relationships between them. By leveraging graph structures to model real-world systems, researchers can systematically analyze the interactions that govern their dynamics. A key challenge in complex network research is identifying important nodes, a topic that has garnered significant attention due to its broad applications. In neural brain networks, neurons are interconnected through nerve fibers, and the transmission of signals enables autonomous decision-making. In transportation networks, identifying critical nodes aids in traffic prediction, optimization, and infrastructure planning. Similarly, in epidemiological networks, infections originating from a few individuals can rapidly propagate through contact links, posing significant risks to public health. In communication networks, detecting key nodes enhances resource allocation and strengthens network security. Likewise, in social networks, pinpointing influential individuals or communities facilitates targeted marketing and improves user engagement. Advances in structural analysis and feature extraction continue to refine our understanding of network robustness, efficiency, and optimization, shedding light on the fundamental principles that govern complex systems.

\begin{table}[htbp]
 \centering
\caption{Comparison of existing studies on critical node identification across different methodological aspects.}
\label{table01}
\renewcommand{\arraystretch}{1} 
\setlength{\tabcolsep}{11pt} 
 \begin{tabular}{lccccccc}
 \toprule
 & \textbf{Centrality} & \textbf{CNDP} & \textbf{IM} &\textbf{Control} & \textbf{AI} & \textbf{Higher-order} & \textbf{Dynamic} \\
 \midrule
 Liu et al.~\cite{REN2013} & \checkmark & \ding{55} & \ding{55} & \ding{55}& \ding{55} & \ding{55} & \ding{55}~\\
 Ren et al.~\cite{xiaolong2014} & \checkmark & \ding{55} & \ding{55} & \ding{55}& 
 \ding{55} & \ding{55} & \ding{55} \\
 Lalou et al.~\cite{lalou2018critical} & \ding{55} & \checkmark & \ding{55} & \ding{55}& \ding{55} & \ding{55} & \ding{55} \\
 Hafiene et al.~\cite{hafiene2020influential} & \ding{55} & \ding{55} & \ding{55}& \checkmark & \ding{55} & \ding{55} & \ding{55} \\
 Zhang et al.~\cite{Zhang2023EpidemicMN} &\ding{55} & \ding{55} & \ding{55}& \ding{55} & \ding{55} & \ding{55} & \ding{55} \\
 Li et al.~\cite{li2023survey} &\ding{55} & \ding{55} & \checkmark & \ding{55}& \ding{55} & \ding{55} & \ding{55} \\
 Jaouadi et al.~\cite{jaouadi2024survey} & \ding{55} & \ding{55} & \checkmark & \ding{55}& \ding{55} &\ding{55} & \checkmark \\
 Liu et al.~\cite{Liu2024Fundamental} & \ding{55} & \ding{55}& \ding{55} & \ding{55}& \ding{55} & \checkmark & \ding{55} \\
 \textbf{Chen et al.(Ours)} & \checkmark & \checkmark & \checkmark & \checkmark & \checkmark &\checkmark & \checkmark \\
 \bottomrule
 \end{tabular}
\end{table}

Recent years have seen a growing number of review articles providing systematic overviews of node importance in complex networks. Various terms—such as key nodes, critical nodes, important nodes, and influence-maximizing nodes—have been used to describe structurally and functionally significant elements, leading to the development of a wide range of methods for assessing node importance in different contexts.
For instance, 
Liu et al.~\cite{REN2013} reviewed advances in node influence ranking by considering both network structure and diffusion dynamics, analyzing the strengths, limitations, and applicability of different approaches. Ren et al.~\cite{xiaolong2014} introduced over 30 methods for identifying important nodes, categorizing them based on criteria such as node neighbors, network paths, feature vectors, and contraction techniques. Their study evaluated node importance from the perspectives of network robustness, vulnerability, and diffusion dynamics, yet the effectiveness of these methods remains constrained by network size and computational complexity. Lalou et al.~\cite{lalou2018critical} focused on the Critical Node Deletion Problem (CNDP), classifying solution algorithms based on problem variants and constraints related to network fragmentation. While their work systematically examined the impact of node removal on network connectivity and vulnerability, its applicability to dynamic and real-world networks remains an open challenge. More recent reviews have explored influence maximization (IM) and node ranking in greater depth. Hafiene et al.~\cite{hafiene2020influential} addressed the IM problem in social networks, aiming to select a set of users that maximize influence spread. However, existing IM approaches often struggle with high computational costs and scalability in large-scale networks. Zhang et al.~\cite{Zhang2023EpidemicMN} classified influential node ranking methods (INRM) and compared validation models based on cascade failure, linear threshold, and epidemic processes, yet integrating these models with real-time applications remains a challenge. Li et al.~\cite{li2023survey} summarized key challenges in diffusion models, such as computational complexity and parameter tuning, and examined how reinforcement learning and graph learning improve efficiency and generalizability. Nevertheless, machine learning–based approaches often require extensive training data and careful model selection. Jaouadi et al.~~\cite{jaouadi2024survey} provided a comprehensive classification of IM models, highlighting their evolution from static to dynamic networks and emerging trends in handling large-scale systems, though trade-offs between accuracy and efficiency remain a concern.
In addition to traditional network models, recent efforts have extended node importance analysis to high-order networks. Liu et al.~\cite{Liu2024Fundamental} reviewed statistical indices for hypergraphs and simplicial complexes, introducing physically meaningful measures such as degree-dependent indicators, clustering coefficients, centrality metrics, and entropy-based indices. 
These studies highlight the diversity of methods for evaluating node importance, as well as significant progress in influence propagation, network robustness, and higher-order structures. However, most previous reviews have focused on specific types of network structures, such as static or higher-order networks, and there is still a lack of comprehensive and unified analyses covering multiple network forms. From the perspective of application scenarios, including influence maximization and network fragmentation, the importance of nodes in relation to network control and stability has been largely overlooked. Similarly, from a methodological perspective, while centrality measures, node deletion strategies, and infection models have been discussed, systematic reviews of machine learning-based approaches and comprehensive index-based methods remain insufficient.  This paper provides a comprehensive review of recent advances in identifying critical nodes in complex networks. Our key contributions are as follows: 

\begin{figure}
    \centering
    \includegraphics[width=0.7\linewidth]{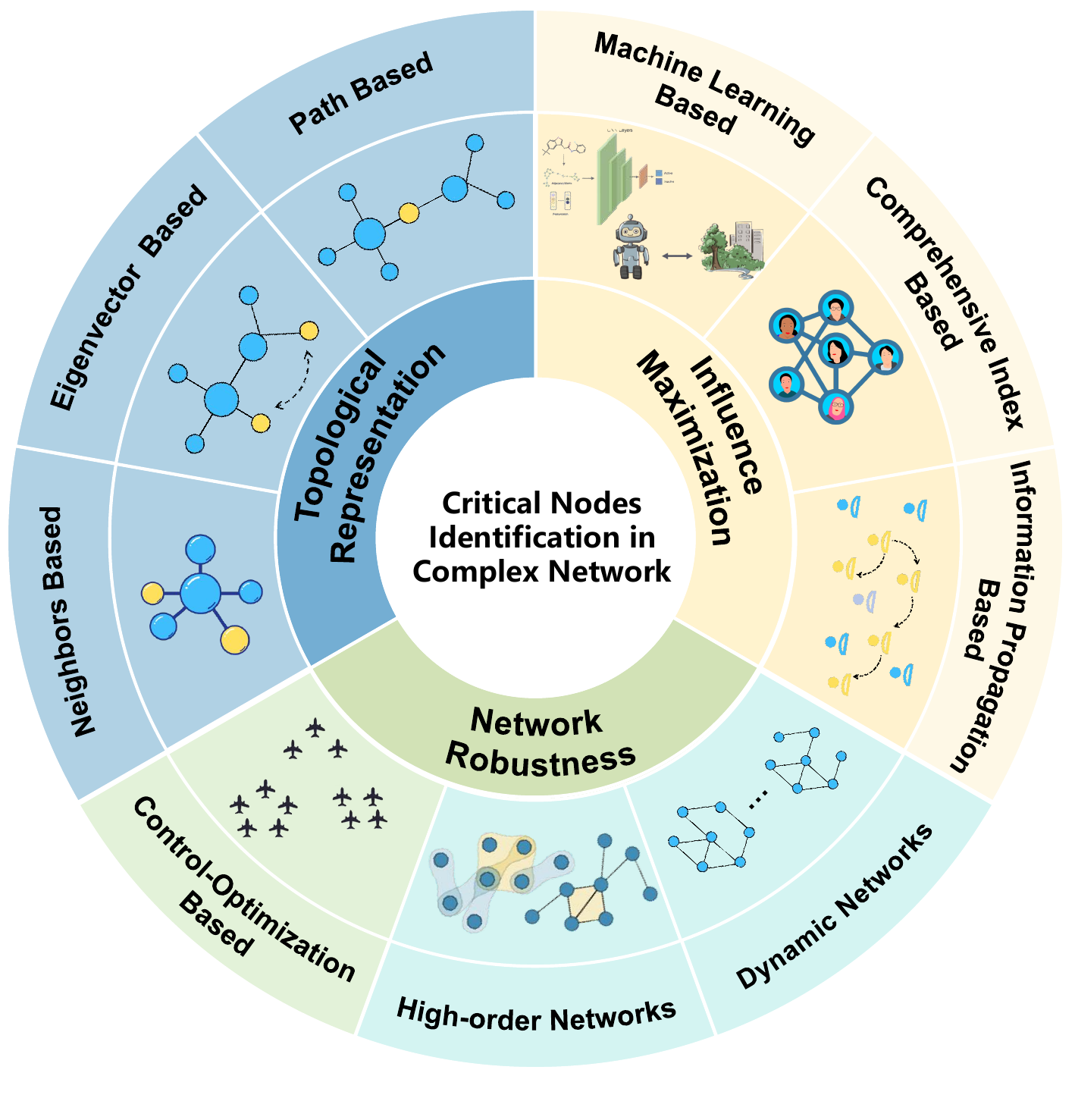}
    \caption{Taxonomy of Critical Node Identification Methods in Complex Networks.}
    \label{fig01}
\end{figure}

\begin{table}[htbp]
\caption{Overview of Structure, Categorization and Contributions.}
\label{table 002}
\renewcommand{\arraystretch}{1.1} 
\begin{tabular}{m{3.5cm}m{5cm}p{7.5cm}} 
 \toprule
\textbf{Ranking Methods} & \textbf{Categories} & \textbf{Details and Contributions}\\
\midrule 
\multirow{6}{*}{Neighbors-Based } & 
\nameref{sec10} & \multirow{6}{=}{Focus on local structural features, including node degrees, coreness, neighbor influence, and cohesive subgraphs, providing intuitive and efficient influence estimation. } \\
 & \nameref{sec11} & \\ 
 & \nameref{sec12} & \\
 & \nameref{sec13} & \\
 & \nameref{sec14} & \\
 & \nameref{sec15} & \\ 
\midrule 
\multirow{4}{*}{Eigenvector-Based} & 
\nameref{sec30} & \multirow{4}{=}{Utilize spectral properties and iterative propagation principles to assess global influence, emphasizing connections to other influential nodes.}  \\
 & \nameref{sec31} & \\
 & \nameref{sec32} & \\
 & \nameref{sec33} & \\
\midrule 
\multirow{4}{*}{Path-Based} 
& \nameref{sec20} & \multirow{4}{=}{Measure node importance based on shortest paths, traversal distances, or probabilistic walks, capturing both local and global topological information.} \\
 & \nameref{sec21} & \\
 & \nameref{sec22} & \\
 &  &  \\
\midrule 
\multirow{4}{*}{Control-Optimization} 
& \nameref{sec43} & \multirow{4}{=}{Formulate node selection as optimization or control problems, identifying key nodes for structural stability, synchronization, or influence maximization.} \\
& \nameref{sec40} & \\ 
 &\nameref{sec41}& \\
 & \nameref{sec42} & \\
\midrule 
\multirow{4}{*}{Machine Learning-Based} & \nameref{sec50} & \multirow{4}{=}{Leverage machine learning models, including classical, deep, and reinforcement learning, to learn influence patterns from structural or dynamic features.} \\
 & \nameref{sec51} & \\
 & \nameref{sec52} & \\
 & \nameref{sec53} & \\
\midrule 
\multirow{4}{*}{Comprehensive Index } & \nameref{sec60} & \multirow{4}{=}{Integrate multiple metrics or criteria into unified indices, achieving more balanced and adaptable node ranking by considering multiple dimensions of importance.} \\
 & \nameref{sec61} & \\
 & \nameref{sec62} & \\
 &  &  \\
\midrule 
\multirow{3}{*}{Information Propagation } & \nameref{sec70} & \multirow{3}{=}{Estimate node influence through simulation or modeling of propagation dynamics, linking rankings to spreading.} \\
 & \nameref{sec71} & \\
 &  &  \\
\midrule 
\multicolumn{2}{l}{High-order Networks} & Capture complex relational patterns beyond pairwise links to improve influence identification accuracy .~\\
\midrule 
\multicolumn{2}{l}{Dynamic Networks} & Address temporal evolution of networks to track time-dependent variations in node importance .~\\
 \toprule 
\end{tabular}
\end{table}

\begin{itemize}
\item \textbf{Categorization of Application Scenarios:} 
From the application scenario oriented perspective of key nodes, this study systematically classifies existing methods into three key categories as depicted in Figure.~\ref{fig01}: (1) topological representation, which assesses node importance by structural metrics such as degree, eigenvalues, and path-based centrality; (2) influence maximization, where the goal is to select a set of Top-$k$ influential seed nodes, $\{v_{1}, v_{2}, \dots, v_{k}\}$, to maximize information spread under diffusion models; and (3) network robustness examines the impact of node removal on network connectivity, modeled as an optimization problem with an objective function $f(G \setminus S)$, where $S \subseteq V$ represents the set of removed nodes. 
\item \textbf{Bridging Gaps in Existing Surveys}: Despite prior studies focusing on specific aspects such as centrality, influence propagation, and critical node deletion problems, there exists a lack of unified discourse regarding the methods and application scenarios of machine learning techniques in recent years. As presented in Table.~\ref{table01}, this study bridges this gap by integrating artificial intelligence methods. These methods will be classified based on their methodological foundations and practical applications. Via comparative analysis, the study underscores the advantages, limitations, and applicability of these methods across different network types. 
\item \textbf{Comprehensive Review of Methodologies:} From a methodological standpoint, this study analyzes nine key node identification techniques across various network structures and dynamic environments as tabulated in Table.~\ref{table 002}. This includes traditional topology-based approaches (e.g., centrality and shortest paths), optimization and control-based frameworks, information diffusion models, machine learning-driven methods, and recent advances in dynamic and higher-order networks. 
\end{itemize}

In this paper, we conduct a structured and comprehensive study of this field. We not only review the existing identification of key nodes in the network and the understanding of application scenarios from the perspective of network structure, methodology, and application scenario, but also discuss key challenges and unresolved issues. Note that the definitions of symbols used in this paper are provided in the Table \ref{table001}. 

\begin{table}[htbp]
 \centering
 \caption{Table of Symbols. 
 }
\label{table001}
 \begin{tabular}{c c }
 \toprule
\textbf{Symbol} & \textbf{Definition} \\
 \midrule
 
\(G\) & Graph \\
\(V = \{v_1, v_2, \dots, v_n\}\) & Set of nodes \\
\(E = \{e_{12}, e_{13}, \dots, e_{ij} | v_i, v_j \in V \}\) & Set of edges, \(e_{ij}\) denotes an edge connecting nodes \(v_i\) and \(v_j\) \\
\(n = |V|\) & Number of nodes \\
\(m = |E|\) & Number of edges \\
\(\Gamma(v)\) & Set of first-order neighbors of node \(v\) \\
\(N(v)\) & Set of first-order and second-order neighbors of node \(v\) \\
\(d_{ij}\) & Path between vertices \(v_i\) and \(v_j\) \\
\(\sigma_{ij}\) & Shortest path length between two nodes \\
\(A = \{a_{ij}\}_{n\times n}\) & Adjacency matrix \\
\(k(v_i)\) & Degree of vertex \(v_i\) \\
\(\Delta_{ijk}\) & Triangle formed by nodes \(v_i\), \(v_j\), and \(v_k\) \\
\(X = [x_1, x_2, \dots, x_n]\) & Eigenvector of the network \\
 \toprule
\end{tabular}
\end{table}

\section{Neighbors-Based Ranking Methods} 
In complex network analysis, the contribution of neighbors can better measure the importance of nodes, and the essence of the node's position and role in the network can be captured through domain information. This section discusses node ranking methods based on the neighbor information of nodes in the network topology as displayed in Table.~\ref{table03}. They can provide a detailed perspective on node importance by considering the local network structure, strike a balance between computational efficiency and accuracy, and are suitable for large-scale networks. 

\subsection{Degree Centrality}\label{sec10}
In the study of social network structures, Bavelas et al.~\cite{RN2} introduced the concept of centrality, and Nieminen et al.~\cite{RN3} addressed the complex and cumbersome calculation of point centrality metrics by proposing a direct measure of node importance through the adjacency matrix . Specifically, the degree of node \( v_i \) is defined as follows: 
\begin{equation}
D'(v_{i}) = \sum_{j=1}^{n} a_{ij}, \quad a_{ij} \in A = \{a_{ij}~\}_{n \times n}.
\end{equation} 
The advantage of this metric is that it is intuitive and easy to understand, with simple computation. However, the node degree metric cannot have comparability across networks of different scales. Building on this, the concept of degree centrality for the entire network is introduced, defined as:
\begin{equation}
k(v_{i})= \sum_{j=1}^{n} a_{ij} /n-1. 
\end{equation} 
The improved metric enables comparisons across networks of varying sizes. Intuitively, a node with more neighbors is likely to be more influential. However, this metric only considers the local state of node \( v_i \) and cannot take into account the local environment of the node. 
To address the limitations of high computational cost and low local relevance in centrality measures, Chen et al.~\cite{RN4} introduced semi-local centrality (SLC), which incorporates the degrees of both first and second-order neighbors with flexible parameters. While degree centrality performs effectively in scenarios with low diffusion probabilities, semi-local centrality is better suited for larger diffusion probabilities, capturing broader network influence. To mitigate the sensitivity of centrality measures to diffusion probabilitie. 
Ma et al.
~\cite{ma2019quasi} proposed Laplacian centrality, which measured node influence by the energy change of a network after node removal. It can be easily computed using node degree, significantly reducing computational complexity and offering good scalability, though it only considered nodes and their first-order information.  Zhu et al.
~\cite{zhu2024identifying} introduced an Improved Laplacian Centrality (ILC) based on the self-consistency concept, incorporating global topological structure information.

\subsection{K-shell Decomposition}\label{sec11}
Kitsak et al.~\cite{RN41} proposed the $k_s$ index based on node local neighbors, which characterizes the topological structure of each node in the network. The method starts by removing all nodes with $k(v)=1$ from the network, and these nodes are defined as $k_s(v)=1$. This process repeats until only nodes with $k(v)\geq 2$ remain. Next, all nodes with $k(v)=2$ in the subgraph are removed, and these nodes are assigned an index of $k_s=2$. This process continues until only nodes with $k(v)\geq 3$ are left. The procedure stops when no nodes remain in the subgraph. In this way, each node is assigned to a corresponding $k$-shell layer, and the overall $k$-shell decomposition is obtained. This method has low time complexity, but it cannot distinguish the importance of nodes within the same layer.

To distinguish the importance of nodes within the same $k$-shell layer, the $k$-shell index overlooks the edges connected to the deleted nodes and the removal order of nodes during the node ranking process. 
Zeng et al.~\cite{RN42} proposed the Mixed Degree Decomposition (MDD), which combines the degree of remaining and removed nodes to better distinguish nodes in the same $k$-shell layer. This method is effective in networks with special structures, such as trees, stars, and regular graphs.
Li et al.~\cite{li2018identification} further classified neighbors by removal order and assigned weights to more accurately capture a node's spreading capability.
Liu et al.~\cite{RN43} estimated node influence using the shortest distance to core nodes, while Zareie et al.~\cite{zareie2018hierarchical} used the distance to the network boundary as a proxy for core proximity. However, the $O(n^2)$ computational complexity of shortest path calculations poses challenges for large-scale networks.

Extensive numerical simulations have shown that in some networks, nodes with high $k$-shell indices, even those classified as core nodes, are not always the most influential in information propagation~\cite{RN43, zareie2018hierarchical}. This indicates a discrepancy between the actual network core and the $k$-shell decomposition index.
Ibnoulouafi et al.~\cite{RN6} introduced M-Centrality, which integrates the $k$-shell index with the degree variation features of both nodes and their neighbors. This metric demonstrates robust performance in dynamic transmission and network connectivity, while maintaining a time complexity of $O(n)$.
Liu et al.~\cite{liu2015core} proposed a link entropy metric that quantifies link diversity within shells by averaging the link diversity of nodes within the same shell. By normalizing for varying node counts, this method effectively distinguishes true core nodes from quasi-core groups. Additionally,
Liu et al.~\cite{liu2015improving} argued that dense interconnections among core nodes could hinder effective information diffusion. They introduced edge diffusion importance by removing edges with low diffusion intensity between nodes, and applying $k$-shell decomposition to the pruned network, yielding a more accurate characterization of node importance.

\subsection{H-index}\label{sec12}
H-index is originally designed to gauge the impact of academic journals and researchers. Lü et al.~\cite{lu2016h} extended the H-index to measure node importance in networks. The H-index of node $v_i$ is defined as:
\begin{equation}
h_{i} = H(k_{j1}, k_{j2}, \dots, k_{jk_{i}}), 
\end{equation}
where $H(\cdot)$ is the operator that finds the largest integer $h$ such that the node has at least $h$ neighbors with degree no less than $h$. If the H-index of node $v_i$ is $h$, it means that node $v_i$ has at least $h$ neighbors with degree no less than $h$.
In large-scale social citation networks, nodes with the same H-index are often hard to distinguish. To address this, Liu et al.~\cite{RN10} proposed the LH-index, incorporating both node and neighbor information. Gao et al.~\cite{RN12} introduced the Weighted H-index for weighted networks by integrating edge weights. Furthermore, Zareie et al.~\cite{zareie2019ehc} proposed the Extended H-index centrality (EHC), which considers the cumulative contributions of all neighbors.

\subsection{K-truss}\label{sec13}
Cohen et al.~\cite{2008Trusses} introduced the concept of K-truss decomposition for a given unweighted undirected network. Let $e_{ij}$ represent the edge between nodes $v_i$ and $v_j$. In the K-truss decomposition, the support of edge $e_{ij}$ is the number of distinct triangles it forms, denoted as:
\begin{equation}
 sup(e_{ij})= \left | {\Delta_{ijk} : \Delta_{ijk} \in \Delta_{G} } \right|, 
\end{equation}
where $\Delta_{ijk}$ represents a triangle with vertices $i$, $j$, and $k$, $k$ is a common neighbor of the two nodes, and $\Delta_{G}$ is the set of all triangles in network $G$. The higher the support of an edge, the higher its clustering characteristic.
The decomposition proceeds based on edge support. Initially, set the parameter to $k=3$. Then, remove edges in the network with support less than $k-2$, and reduce the support of the two other edges forming a triangle with each edge by 1, updating the network topology. Continue removing edges with support less than $k-2$ until all edges in the network have support greater than or equal to $k-2$. Nodes removed in this stage are assigned to the k-truss layer. For the next step, increase $k+1$, and repeat the process of removing edges with support less than $k-2$ until all nodes in the network are completely assigned to the corresponding layers.
Malliaros et al.~\cite{malliaros2016locating} introduced a triangle-graph-based k-core decomposition that refines the analysis of densely connected networks. Their method calculates the "trustness" of an edge, defined by its participation in a k-truss (the number of triangles it is part of). This approach offers a more detailed analysis compared to $k$-shell decomposition, making it particularly effective for identifying influential communities within networks.

\subsection{Structural Hole}\label{sec14}
Yu et al.~\cite{yu2017critical} introduced the concept of "structural hole" to identify key nodes in a network by defining a node influence matrix $P$. If there is no direct connection or indirect redundancy relationship between two entities or communities, the obstruction between them is called a structural hole:
\begin{equation}
 p_{ij} = \frac{a_{ij}}{\sum_{j \in \Gamma(v_i)} a_{ij}}, 
\end{equation}
where $p_{ij}$ represents the proportion of total effort that node $v_i$ contributes to maintaining node $v_j$ as a neighbor, and $\Gamma(v_i)$ is the set of neighbors of node $v_i$. The network constraint coefficient of node $v_i$ is calculated as:
\begin{equation}
	C_{i} = \sum_{j \in \Gamma(v_i)} \left( p_{ij} + \sum_{q} p_{iq} p_{qj} \right)^{2}, \quad q \neq i, j, 
\end{equation}
where $q$ is an intermediate node connecting $v_i$ and $v_j$, and $p_{iq}$ and $p_{qj}$ represent the total effort proportions of nodes $v_i$ and $v_j$ in maintaining their neighborhoods with a common neighbor $q$. A smaller constraint coefficient signifies a larger structural hole, making the node more influential in information propagation.
Yu et al.~\cite{yu2017identifying} proposed an improved structural hole method (ISH) that integrates degree and nearest-neighbor information, effectively identifying critical nodes for information propagation by considering edge weights based on neighbor degrees.

In contrast, while the $k$-shell index prioritizes central nodes, it overlooks neighbor interactions and broader topological features. To address this, Xu et al.~\cite{xu2018identifying} introduced a local centrality measure (VKC) that combines neighborhood information with structural hole concepts, using outward-link diversity and network constraint coefficients.
Building on these ideas, Liu et al.~\cite{liu2021identifying54} developed an information control capability index (ICC), integrating $k$-shell values and enhanced structural hole theory. Zhao et al.~\cite{zhao2023ranking21} further refined this approach with the SHKS node ranking index, combining $k$-shell and structural hole metrics, identifying vital bridging nodes with lower $k$-shell indices.
While "structural holes" focus on nodes controlling parts of the network, they typically overlook the broader network topology. The combination of structural hole theory and $k$-shell methods enhances the overall network capacity~\cite{ma2023effective}.

\subsection{Neighborhood Similarity}\label{sec15}
In a network, two similar nodes are more likely to be interested in transmitting the same information, and the likelihood of information propagating through the edges between the nodes is higher. Wang et al.~\cite{wang2015link} defined the similarity between nodes $v_i$ and $v_j$ based on the number of their common neighbors. Specifically, the more common neighbors two nodes have, the more similar they are. The similarity is mathematically defined as:
\begin{equation}
 HP(v_{i}, v_{j}) = \frac{\left | \Gamma(v_{i})\cap \Gamma(v(j)) \right | }{\min (|\Gamma(v_{i})|, | \Gamma(v(j)|)}, 
\end{equation}
where $v_{i}$ and $v_{j}$ represent the starting node and the randomly selected neighboring node, respectively. 
Lu et al.~\cite{lu2022critical11}  proposed an Extended Similarity Coefficient Ranking Method (ESCRM), which connected the importance of nodes to the local structural similarity between nodes and their neighbors. Ai et al.~\cite{ai2023identifying34} introduced Resource Allocation Similarity (RAS), simulating resource transfers based on the inverse sum of common neighbors to capture both structural and functional similarities.  Rao et al.~\cite{rao2022cbim} decovered small communities from the multilayer network by dice neighborhood similarity~(DNS), merging smaller communities and generating larger communities to maximize community influence and find k seed nodes in the multilayer network. Tong et al.
~\cite{tong2024novel} proposed an Extended Tanimoto Correlation~(TC) for topological clustering of neighborhood sets to measure node importance.

\begin{table}[htbp]
 \centering
 \caption{Comparison of Neighbors-Based Ranking Methods. 
 }
\label{table03}
 \begin{tabular}{m{4cm}m{4cm}m{7cm}}
 \toprule
 Title & Related Works & Advantages / Disadvantages\\
 \midrule
 Degree Centrality &~\cite{RN2},~\cite{RN3},~\cite{RN4}, ~\cite{ma2019quasi},~\cite{zhu2024identifying} & 
 $+$ Simple and computationally efficient. 
 
 $-$ Fails to capture global network structure, only reflects local connectivity of neighbours.  \\
 \midrule
 $k$-shell Index &~\cite{RN41},~\cite{RN42},
~\cite{li2018identification}, 
~\cite{RN43}, 
~\cite{zareie2018hierarchical}, 
~\cite{RN6}, 
~\cite{liu2015core}, 
~\cite{liu2015improving} 
 &
 $+$ Suitable for large-scale networks, identifies core nodes. 
 
 $-$ Cannot differentiate nodes within the same layer, information spread may not correlate directly with $k$-shell values. \\ 
 \midrule
 H-index &\cite{lu2016h},~\cite{RN10},~\cite{RN12},~\cite{zareie2019ehc} & 
 $+$ Incorporates degree and neighbor influence.
 
 $-$ Cannot distinguish nodes with the same H-index, dependent on neighborhood degree distribution. \\
 \midrule
 K-truss & 
~\cite{2008Trusses},~\cite{malliaros2016locating} & 
 $+$ Identifies community cores and modules. 
 
 $-$ High computational cost, ineffective in sparse networks. \\
 \midrule
 Structural Hole & 
~\cite{yu2017critical},~\cite{yu2017identifying},~\cite{xu2018identifying},~\cite{liu2021identifying54},~\cite{zhao2023ranking21},~\cite{ma2023effective} & 
 $+$ Identifies bridging nodes, effective for information spread.
 
 $-$ Computationally expensive, requires global information. \\
 \midrule
 Neighborhood Similarity & \cite{wang2015link},~\cite{lu2022critical11},~\cite{ai2023identifying34} ,~\cite{rao2022cbim},~\cite{tong2024novel}& 
 $+$ Beneficial for social networks and recommendation systems. 
 
 $-$ High computational complexity in large networks. \\
 \bottomrule
 \end{tabular}
\end{table}

\section{Eigenvector-Based Ranking Methods}
Eigenvector-based methods evaluate node importance by considering both the number and influence of connections through eigenvector analysis and iterative computation as details in Table.~\ref{table04}. These methods extend beyond simple neighbor counting by incorporating the significance of connected nodes. They are extensively used in social network analysis to identify influential entities, in web search algorithms for page ranking, and in biological networks to detect critical components. The key strength of these methods lies in their ability to capture the recursive nature of importance: a node is deemed important if it is connected to other important nodes. This property enables more accurate assessment of node significance compared to approaches based solely on local neighborhood information. Nevertheless, such methods entail several limitations, including high computational cost, sensitivity to parameter settings, and the necessity for cautious interpretation across diverse network contexts. 

\subsection{Eigenvector Centrality}\label{sec30}
Eigenvector Centrality (EC) takes into account the importance of a node's neighbors~\cite{bonacich1971factoring}~\cite{bonacich2007some}. It uses the eigenvectors and eigenvalues of the network's adjacency matrix to measure the importance of nodes. The main idea is to assume that the importance of node $v_i$ is denoted as $x_i$. Thus, the eigenvector centrality of node $v_i$ is defined as:
\begin{equation}
EC(v_{i}) = x_{i} = c \sum _{j=1}^{n} a_{ij}x_{j}, 
\end{equation}
where $c$ is the proportionality constant, and initializing $x = [x_{1}, x_{2}, \dots, x_{n}]$, after several iterations, it converges to $x = cAx$, rewritten as $Ax = c^{-1}x$, meaning $x$ is the eigenvector corresponding to the eigenvalue $c^{-1}$ of the adjacency matrix $A$. 
Building on this, Ilyas et al.~\cite{ilyas2011identifying} introduced Principal Component Centrality (PCC) to identify social hubs in large-scale networks, measuring node influence based on its position using eigenvectors of the adjacency matrix.
To identify influential members in social networks, Estrada et al.~\cite{PhysRevE71} defined subgraph centrality, which sums closed paths within a node's subgraph, with shorter paths contributing more. Ahajjam et al.~\cite{ahajjam2015leadersrank} proposed the LeadersRank algorithm, ranking nodes based on eigenvector centrality and identifying communities through second-order neighbors. However, this algorithm primarily bases node importance on immediate neighbors, leading to slow convergence and neglect of external influences.

To address eigenvector centrality's localization issue, which concentrates weight on a few nodes, Martin et al.~\cite{martin2014localization} introduced a method based on a non-backtracking matrix. This method avoids localization while yielding similar results to eigenvector centrality in dense networks.  Zhong et al.~\cite{zhong2018identifying}  combined Jaccard similarity with the dissimilarity between nodes and proposed the ECDS centrality method, which takes the weighted sum of the centralities of its neighboring nodes to form an iterative formula similar to the eigenvector centrality.

\subsection{Katz Centrality}\label{sec31}
Katz centrality~\cite{katz1953new} extended eigenvector centrality by adding a constant term to describe the importance of the central node itself. It is defined as:
\begin{equation}
 x_{i} = c\sum_{j=1}^{N} A_{i, j}x_{j} + \beta, 
\end{equation}
where $\beta$ is a constant, and in matrix form, it can be represented as $x = cAx + \beta$, which is equivalent to $x = (I - cA)^{-1}~\beta$. Here, $x$ represents the Katz centrality vector for all nodes in the network. It assigns different weights to all paths in the network, assuming that longer paths are less important than shorter ones. However, it has a high time complexity and is only applicable to smaller networks with few loops.
To address the limitation of betweenness centrality, which only considers the shortest paths and ignores non-shortest paths, Zhang et al.~\cite{zhang2015identifying} proposed the BKC algorithm, which combines path-information and Katz centrality. This approach takes into account both local and global features, overcoming the limitation of evaluating node importance based solely on neighboring nodes.

\subsection{Pagerank Algorithm}\label{sec32}
Traditional ranking methods for webpages rely on "keyword density, " but this can be undermined by "malicious keywords." Page et al.~\cite{page1998pagerank} proposed the core algorithm of Google's search engine, Pagerank. The value of node $v_i$ at time $t$ is defined as:
\begin{equation}
 PR_{i}(t) = \sum_{j=1}{n} a_{ji} \frac{PR_{j}(t-1)}{k_{j}^{out}}, 
\end{equation}
where $k_{j}^{out}$ represents the out-degree of node $v_j$. However, in practice, there are cases where webpages are unreachable. To address this issue, a random jump probability $c$ is introduced:
\begin{equation}
 PR_{i}(t)= (1-c)\sum_{j}^{n} a_{ji} PR_{j}(t-1) / {k_{j}^{out}} + c/n, 
\end{equation}
where $c$ is a constant derived from experiments. In the PageRank algorithm, each node has the same random jump probability, and the parameter $c$ can be flexibly adjusted based on the network characteristics. The algorithm measures the importance of a webpage based on the quality and quantity of its inbound links. If a webpage is linked by many pages or high-quality pages, it is considered to have greater importance.

To address the limitations of PageRank, such as its failure to incorporate node attributes or external information, Yang et al.~\cite{yang2017mining} described the frequency and duration of contact between nodes in a directed weighted network to capture the importance of nodes based on their in-degrees and out-degrees. They proposed the Two-Way-PageRank method to identify important nodes in directed weighted networks.
Hsu et al.~\cite{hsu2017unsupervised} introduced the AttriRank model, which integrates node attribute similarity into the ranking process. This model utilizes random walks, capturing associations between nodes based on both network structure and attribute similarity. The random walk is framed as a Markov chain, with a transition matrix that balances information propagation across the network and node attributes.
Building on this, Sheng et al.~\cite{sheng2020identifying12} proposed the Trust-PageRank (TPR) algorithm, which combines attribute similarity and trust values, defined through structural degree ratios. TPR effectively merges node attributes and network structure to refine node rankings.
Su et al.~\cite{su2021identification67} introduced a cascade failure and directionally weighted PageRank method tailored for power grid networks. This approach incorporates grid connectivity, electrical topology, and power flow direction, thus enhancing the grid’s resilience by mitigating the impact of cascading failures.

\begin{table}[htbp]
 \centering

 \caption{Comparison of Eigenvector-Based Ranking Methods.}
\label{table04}
 \begin{tabular}{m{4cm}m{4cm}m{7cm}}
 \toprule
 Title & Related Works & Advantages / Disadvantages \\
 \midrule
 Eigenvector Centrality & 
~\cite{bonacich1971factoring},~\cite{bonacich2007some},~\cite{ilyas2011identifying}, ~\cite{PhysRevE71} ,
~\cite{ahajjam2015leadersrank},~\cite{martin2014localization},~\cite{zhong2018identifying} 
 & 
 $+$ Considers neighbor importance, widely used in networks.
 
 $-$ Suffers from localization issues, slow convergence. \\
 \midrule
 Katz Centrality & 
~\cite{katz1953new},~\cite{zhang2015identifying} 
 & 
 $+$ Balances local and global influence. 
 
 $-$ High computational cost, sensitive to parameters. \\
 \midrule
 PageRank Algorithm & 
~\cite{page1998pagerank},~\cite{yang2017mining},~\cite{hsu2017unsupervised}, 
~\cite{sheng2020identifying12},~\cite{su2021identification67} 
 & 
 $+$ Scales well for large networks, robust.
 
 $-$ Sensitive to damping factor, ignores node attributes. \\
 \midrule
 Voting Mechanism & 
~\cite{RN33}, 
~\cite{li2014identifying}, 
~\cite{RN36},~\cite{RN38}, 
~\cite{RN37}, 
~\cite{li2022improved23}, 
~\cite{liu2016evaluating} 
& 
 $+$ Simple and adaptable for various networks. 
 
 $-$ Affected by network structure, struggles in dense graphs. \\
 \bottomrule
 \end{tabular}
\end{table}

\subsection{Voting Mechanism}\label{sec33}
To address the parameter adjustment issue in the PageRank algorithm, Lü et al.~\cite{RN33} introduced a background node $g$ in the network, which establishes bidirectional connections with all other nodes, making the network strongly connected. They proposed the LeaderRank~(LR) algorithm. Initially, all pages are assumed to have a LR value of 1, while the background node has a LR value of 0. After convergence, the LR value for node $i$ is defined as:
\begin{equation}
 LR_{i} = LR_{i}(t_{c})+ \frac{LR_{g}(t_{c})}{n}.
\end{equation}
After traversing all the nodes in the network, the LR value is obtained. Compared to PageRank, LeaderRank has stronger robustness in resisting spammer attacks and random interference. Building on this, Li et al.~\cite{li2014identifying} introduced the weight-LeaderRank algorithm for weighted networks, which modifies the standard random walk by incorporating a biased random walk. This adjustment ensures that neighboring nodes receive higher scores, thereby improving the identification of influential propagators. 
To address the issue of overlapping influence in methods such as PageRank, ClusterRank, and $k$-shell, Zhang et al.~\cite{RN36} proposed the iterative VoteRank method. In this approach, nodes receive votes from their neighbors, and their importance is determined by the number of votes they accumulate. This method reduces to degree-based ranking when all nodes have the same voting ability.

Building on the information of node neighborhood within the network topology, Kumar et al.~\cite{RN38} proposed the NCVoteRank method, which adjusts the voting weights according to the core degree of neighboring nodes, adding flexibility to the voting framework. Sun et al.~\cite{RN37} enhanced this concept by introducing the weight-VoteRank method, which incorporates both the number of neighbors and edge weights, allowing for a more nuanced measure of node importance in weighted networks. Li et al.~\cite{li2022improved23} further refined the VoteRank algorithm with the introduction of Linearity and Importance (DIL)~\cite{liu2016evaluating}. This method initializes node scores and voting abilities based on local importance, and updates the voting capabilities of neighbors in each round. Through iterative voting, the Top-$k$ nodes with the highest scores are selected, improving the efficiency and accuracy of the ranking process. These methods collectively advance the VoteRank algorithm by integrating node neighborhood information, edge weights, and local importance, thereby enhancing the precision and efficiency of node ranking in complex networks.

\section{Path-Based Ranking Methods}
Path-based ranking methods assess node importance through the analysis of paths connecting nodes, taking into account their positions and roles within network connectivity and information dissemination. These methods characterize the global structure and connection patterns of networks by evaluating shortest paths, average distances, and random walk processes as details in Table.~\ref{table05}. Applications include identifying intermediaries and pivotal communicators in social networks, locating critical hubs in transportation systems, and understanding material or information transfer mechanisms in biological networks. The primary advantage of these approaches lies in their capability to capture the global influence of nodes and highlight those that function as bridges or central connection points within the network. However, these methods are computationally intensive, particularly for large-scale networks, necessitating simplifications or approximations to ensure practical feasibility. 

\subsection{Betweenness Centrality}\label{sec20}
Freeman et al.~\cite{RN56} reviewed traditional centrality measures to determine the importance of nodes in social networks. Based on the connectivity between nodes, they introduced Betweenness Centrality (BC), which measures the importance of a node by counting the number of shortest paths that pass through the node. It is defined as:
\begin{equation}
 BC(v)= \sum_{i, j\ne v}~\frac{\left | d_{ij}(v) \right |} {\left | d_{ij} \right |}, 
\end{equation}
where $ \left | d_{ij} \right | $ is the number of shortest paths between nodes $v_i$ and $v_j$, and $ \left | d_{ij}(v)\right | $ is the number of shortest paths between nodes $v_i$ and $v_j$ that pass through node $v$. BC takes into account the global features of the network, but its path computation has a high time complexity. Hage et al.~\cite{hage1995eccentricity} proposed using the maximum distance between a node and all other nodes in the network as a measure of the node's eccentricity centrality, $\text{ECC}$. The eccentricity centrality of node $v_i$ is defined as:
\begin{equation}
 \text{ECC}(v_{i}) = \max_{j}(d_{ij}), \quad j=1, 2, \dots, n, 
\end{equation}
where the network diameter is the maximum eccentricity value among all nodes in the network $G$, and the network radius is defined as the minimum eccentricity value among all nodes. Clearly, the node whose eccentricity centrality value equals the network radius is the central node. Boccaletti et al.~\cite{RN59} proposed the ASP index, which measures a node's importance based on its effective information transfer to all other nodes. The ASP index is defined as the sum of the shortest path distances from a node to all other nodes, standardized by the total number of node pairs in the network. However, the ASP index becomes ineffective when a node is removed, as it can disconnect the network.To address this, Lü et al.~\cite{RN61} introduced the Relative Change of Average Shortest Path (RASP) centrality, which focuses on the change in the average shortest path in the network after a node is removed. RASP measures the node's importance in maintaining network connectivity by calculating the relative change in the average shortest path.

To address the high time complexity of betweenness centrality, Song et al.~\cite{song2015novel} introduced load centrality (LC), which measures the change in network paths after node removal to assess the node's impact on connectivity.
Ventresca et al.~\cite{ventresca2015efficiently} developed the CNICC algorithm, which determines node connectivity and capacity based on path length, the number of paths, and network scale. 
Many real-world networks display hierarchical and modular structures. Zhang et al.~\cite{zhang2011node} proposed a multi-scale node importance metric using a kernel function, where bandwidth determines the range of interactions. Small bandwidth captures short-range interactions, while large bandwidth highlights long-range interactions, offering insights into node influence across different scales during dynamic processes.
In power systems, Yang et al.~\cite{yang2020critical} introduced the Electric Observability Capability Index (EOCI) and the Electrical Dynamic Characteristics Index (EkI) to identify critical nodes influencing power system observability and controllability.
Kianian et al.~\cite{kianian2021efficient} used degrees and independent influence paths to approximate influence propagation, reducing computational cost by pruning insignificant nodes. Xiao et al.~\cite{xiao2024new} applied LASPN theory to identify critical nodes, combining local average shortest paths and extended neighborhoods for efficient processing in large-scale networks.

\subsection{Closeness Centrality}\label{sec21}

To address the interference of special values in paths, Freeman et al.~\cite{freeman2002centrality} introduced Closeness Centrality by calculating the average distance of a node to all other nodes in the network. The smaller the average shortest distance $d_i$ from node $v_i$ to other nodes in the network, the closer node $v_i$ is to the other nodes in the network. The reciprocal of $d_i$ is defined as the closeness centrality of node $v_i$:
\begin{equation}
 CC(i)=\frac{1}{d_i} = \frac{n-1}{\sum_{j\ne i}d_{ij}}.
\end{equation}
In a connected network, the smaller the average distance between a node and all other nodes in the network, the larger the closeness centrality of that node. This can be understood as determining the importance of a node by the average transmission time of information in the network.
To improve closeness centrality for network efficiency, Latora et al.~\cite{latora2001efficient} extended betweenness centrality to disconnected networks and proposed the EFF index. However, this approach requires calculating the distances between all node pairs, which is computationally expensive for large networks.
To reduce this complexity, Salavati et al.~\cite{salavati2019ranking} applied the Louvain community detection algorithm to identify network communities. By maximizing modularity, they extracted community structures and selected key nodes using betweenness centrality. They also introduced a method to identify gateway nodes linking communities and reduced computation by focusing on a subset of special nodes.
Okamoto et al.~\cite{Okamoto7328} proposed the RAND algorithm to estimate average node distances efficiently. It selects top candidate nodes based on these estimates and ranks them precisely, balancing computational efficiency with accuracy.
Sheng et al.~\cite{sheng2020identifying} introduced the GLS method, which calculates the influence of nodes by considering common neighbors and structural characteristics. This method combines local and global structural influence based on their information exchange capabilities.

\begin{table}[htbp]
 \centering
 \caption{Comparison of Path-Based Ranking Methods.}
\label{table05}
 \begin{tabular}{m{4cm}m{4cm}m{7cm}}
 \toprule
 Title & Related Works & Advantages / Disadvantages\\
 \midrule
 Betweenness Centrality & 
\cite{RN56}, 
~\cite{hage1995eccentricity}, 
~\cite{RN59}, 
~\cite{RN61}, 
~\cite{song2015novel}, 
~\cite{ventresca2015efficiently}, 
~\cite{zhang2011node}, 
~\cite{yang2020critical}, 
~\cite{kianian2021efficient}, 
~\cite{xiao2024new} 
 & 
 $+$ Captures shortest paths, global influence, robustness.
 
 $-$ High computational cost, sensitive to network changes. \\
 \midrule
 Closeness Centrality & 
~\cite{freeman2002centrality},~\cite{latora2001efficient},~\cite{salavati2019ranking},~\cite{Okamoto7328},~\cite{sheng2020identifying} 
 & 
 $+$ Measures reachability efficiency, effectiveness.
 
 $-$ Ineffective in disconnected networks, computationally expensive in large graphs. \\
 \midrule
 Random Walk & 
~\cite{iannelli2018influencers},~\cite{RN7},~\cite{kermarrec2011second} 
 & 
 $+$ Models information spread, considers network topology, improves influence ranking. 
 
 $-$ Computationally intensive, rely on walk length. \\
 \bottomrule
 \end{tabular}
\end{table}

\subsection{Random Walk}\label{sec22}
Currently, methods for identifying key nodes based on heuristic centrality measures are used only for specific network topologies or unique dynamic models for analysis and demonstration. 
Iannelli et al.~\cite{iannelli2018influencers} proposed a method based on the random walk effective distance $D^{RW}_{ij}(\lambda)$ between pairs of nodes, using the adjacency matrix $A_{ij}$ and the diffusion dynamic parameter $\lambda$ to quantify the influence of nodes during the network diffusion process. The ViralRank score for node $v_i$ is defined as the average random walk effective distance between all source nodes and target nodes in the network:
\begin{equation}
\begin{aligned}
 & v_{i}(\lambda) = \frac{1}{N} \sum_{j} \left( D_{ij}^{RW}(\lambda) + D_{ji}^{RW}(\lambda) \right), \\
 & D_{ij}^{RW}(\lambda) = - \text{In}~\sum_{k\ne j} \left( I^{j} - e^{-\lambda}P^{(j)} \right )^{-1}_{ik} e^{-\lambda}p_{k}^{(j)}, 
\end{aligned}
\end{equation}
where $i \neq j$, $D_{ii}^{RW} = 0$, $P^{(j)}, I^{(j)} \in R^{(N-1)\times (N-1)}$ are partition matrices of the Markov matrix, $(P)_{ij} = A_{ij} / \sum_{k} A_{ik}$, and the identity matrix $(I)_{ij} = \theta_{ij}$ are obtained by removing the $i$-th row and $j$-th column, respectively. $p^{(j)}$ is the $j$-th row of $P$ after removing the $v_j$-th element. A node's smaller average effective distance means a higher ViralRank score, indicating central nodes cause larger outbreaks than peripheral ones as "seed" nodes.
Dong et al.~\cite{RN7} proposed the Semi-Local Centrality~(SLC) method, integrating topological features with dynamic processes. SLC starts a random walk from node $v_i$, adding influential neighboring nodes to the path. The semi-local centrality of a node is defined by the length of the random walk path formed by its influential neighbors, improving the accuracy.
To address the limitations of current centrality measures and their incompatibility with distributed environments, Kermarrec et al.~\cite{kermarrec2011second} proposed second-order centrality, which calculates the standard deviation of return times recorded by each node during a random walk. This approach supports distributed computation and provides a local value for each node, reflecting its relative importance. It allows unbiased identification of key nodes through random walk, enabling a global understanding of network topology.

\section{Control-Optimization Based Ranking Methods}
Control optimization-based ranking methods identify key nodes by employing network control theory and optimization techniques. These methods aim to characterize the influence of nodes on network dynamics, controllability, and connectivity, offering valuable insights for applications in security, robustness enhancement, and regulation of dynamic processes as details in Table.~\ref{table06}. In security, they help detect vulnerable links and improve resilience against attacks. In infrastructure networks, they aid in designing effective control strategies. The primary advantage of these approaches lies in their ability to offer a dynamic and interactive perspective on node importance, revealing the impact of individual nodes on overall network functionality and stability. However, these methods face challenges such as high computational complexity as well as control and optimization theory.

\subsection{Network Control}\label{sec43}
Existing research has highlighted the importance of specific nodes in regulating network dynamics and functionality. Li et al.~\cite{li2006controlling} employed the Watts-Strogatz propagation model to explore spreading mechanisms in small-world networks. Their analysis of delayed control propagation with linear and nonlinear feedback controllers, demonstrating how network parameters influence stability and oscillatory behavior. 
Ghosh et al.~\cite{ghosh2022synchronized} emphasized the critical role of dynamic synchronization in maintaining global network coherence. Their findings showed that proper weighting procedures significantly enhance synchronization in static complex networks, with key nodes exerting disproportionate influence on system stability.
D'Souza et al.~\cite{d2023controlling} proposed a framework for network control that integrates node dynamics and macroscopic network properties, refining control strategies through higher-order interactions. 
Liu et al.~\cite{liu2011controllability} developed analytical tools to assess the controllability of arbitrary directed networks, establishing that the number of driver nodes required for full control is primarily dictated by the network's degree distribution. Their work uncovered a fundamental relationship between network structure and control feasibility: sparse, heterogeneous networks are the most difficult to control, whereas dense, homogeneous networks require fewer driver nodes. The controllability of a system governed by linear time-invariant dynamics:
\begin{equation}
 \frac{d x(t)}{dt} = Ax(t) + Bu(t), 
 \end{equation}
where $x(t)\in R^{N}$  represents the system state of $N$ nodes at time $t$, $A\in R^{N\times N}$ is the adjacency matrix encoding interaction dynamics, and $B\in R^{N\times M}$ identifies the driver nodes influenced by control inputs
$u(t)\in R^{M}$. The system is controllable if and only if the controllability matrix, $
 C = (B, AB, \dots, A^{n-1}B) \in \mathbb{R}^{N \times NM}$.
The system is controllable if and only if $Rank(C)=N$, meaning it can transition from any initial state to any final state within a finite time. 
Furthermore, the minimum number of driver nodes $N_D$ required for full control can be estimated using the structural controllability framework as
$
 N_D = \max\{1, N- \operatorname{rank}(A)\}.
$
To address the problem of identifying an optimal driver node set for minimizing control cost, recent studies have proposed formulations such as
\begin{equation}
\min_{B} |u(t)|^2 \quad \text{subject to} \quad \frac{d x(t)}{dt} = Ax(t) + Bu(t),
\end{equation}
which seeks the control input $u(t)$ with the smallest energy expenditure required to guide the system toward the desired state. 
To address the minimal-cost control problem, 
Ding et al.~\cite{ding2017key} proposed the Projection Gradient Method Extension (R-PGME) algorithm, which leverages Monte Carlo simulations and identifies a minimal set of key nodes for effective control. Lu et al.~\cite{lu2019pinning} applied the Warshall algorithm to Boolean control networks (BCNs), determining the smallest set of nodes required for stability in large-scale systems. 
Expanding on these approaches, Zhu et al.~\cite{Zhu10106394} integrated structural information and developed a low-complexity structural controller for large-scale Boolean networks, efficiently identifying the minimal feedback node set for robust control.

The eigenvalues of the Laplacian matrix play a critical role in pinning control and network controllability, directly influencing the selection of optimal driver nodes for synchronizing large-scale networks. Yu et al.~\cite{yu2012distributed} proposed a distributed adaptive strategy that adjusts coupling weights based on local node dynamics to achieve synchronization, later extending their work to establish synchronization criteria for various network topologies, including strongly and weakly connected graphs~\cite{yu2013synchronization}. Ding et al.~\cite{ding2016optimizing} demonstrated that denser and more homogeneous networks exhibit superior controllability, with controlled nodes typically having lower in-degree than the network average.
Amani et al.~\cite{amani2017finding} introduced the Eigenvalue Sensitivity Index (ESI), which ranks nodes using extremal Laplacian eigenvectors to estimate their importance in pinning control. They further developed a computationally efficient method based on sensitivity analysis of the Laplacian matrix to identify near-optimal driver nodes~\cite{amani2018controllability}. Liu et al.~\cite{liu2018optimizing} evaluated pinning effectiveness via the smallest eigenvalue of the grounded Laplacian, linking controllability to network structure. 
Several optimization strategies have been investigated. Wang et al.~\cite{wang2012optimizing} achieved full network control through structural perturbations using a single driver node. Bof et al.~\cite{bof2016role} showed that energy-efficient control depends on the left and right Perron eigenvectors of the network matrix. Zhou et al.~\cite{zhou2018node} proposed the ControlRank metric, validating degree-based rankings in scale-free networks. Liu et al.~\cite{hui2021node} refined driver node selection by prioritizing nodes with the largest eigenvector components corresponding to the dominant Laplacian eigenvalue. Bomela et al.~\cite{bomela2023finding} applied the Moore–Penrose pseudoinverse of the Laplacian to identify Most Influential Nodes (MIN) in oscillatory networks.
Recent studies have emphasized minimal intervention strategies. Jiang et al.~\cite{jiang2024pinning} showed that large-scale synchronization can often be achieved with minimal control effort or higher-order interactions. Sun et al.~\cite{sun2025optimal} analyzed optimal control in reaction-diffusion networks, revealing the influence of topology and diffusion parameters on epidemic dynamics.
Collectively, these works highlight the strong interdependence between spectral properties, network structure, and dynamic processes, underscoring the value of targeted interventions for achieving efficient and robust network control.

The contribution of network edges, including loops, to overall network dynamics is fundamentally governed by the eigenvector corresponding to the smallest nonzero eigenvalue of the Laplacian matrix, known as the Fiedler vector. The cumulative effect of all edges is encapsulated in the network’s algebraic connectivity, which dictates synchronization and diffusion properties.
Zhang et al.~\cite{zhang2017effect} analyzed the impact of adding feedback edges in directed acyclic graphs (DAGs) on consensus behavior in multi-agent systems. They identified disruptive feedback edges that degrade convergence speed and formulated a necessary and sufficient condition for their identification. Mo et al.~\cite{mo2019effects} extended this by introducing a topology-based optimization framework for directed graphs, enabling the systematic selection of edges that enhance convergence. They demonstrated that in strongly connected weighted digraphs, the second smallest Laplacian eigenvalue remains real and established conditions for edge selection that improve consensus dynamics.
Jiang et al.~\cite{jiang2023searching} proposed a method to assess the importance of loops in complex networks, ranking key loops based on their influence on the Fiedler value. Similarly, Cao et al.~\cite{cao2024synchronization} developed a theoretical framework for synchronization acceleration by optimizing directed edge placement and weights. They derived a necessary and sufficient condition for selecting edges that enhance convergence, allowing efficient identification through Laplacian eigenspace computations.
Zhang et al.~\cite{zhang2022necessary} reaffirmed the existence of disruptive feedback edges in DAGs and explicitly constructed conditions for their identification. Gao et al.~\cite{gao2025effects} further examined how weighted edge additions influence algebraic connectivity in directed graphs, proving that changes in Laplacian eigenvalues are localized within subgraphs containing the added edges. Their findings underscore the dependence of algebraic connectivity on edge weight, range, and distribution along network paths. These studies further reveal how interventions of network edges and cycles control dynamic behavior, highlighting the role of spectral properties in optimizing synchronization and controllability.

\subsection{Maximization of Connected Component}\label{sec40}
In a network, there exists a set of important nodes that influence the network's robustness and connectivity. By removing these nodes or attacking the network, we can identify a critical set of nodes. Deleting this set of nodes can minimize or maximize the connectivity components (such as the largest connected component or node pair connectivity) in the remaining network, which is known as the Critical Nodes Deletion Problem.
 Aringhieri et al.~\cite{aringhieri2016general} designed a general framework to address several classic critical node problems (CNPs), specifically: (1) \textbf{Pairwise connectivity}: Minimize the number of connected node pairs in the remaining network by deleting up to $K$ nodes, ensuring the pairwise connectivity is reduced to a threshold $P$.
(2) \textbf{Largest connected component}: 
Minimize the size of the largest connected component in the remaining network by deleting up to $K$ nodes, ensuring its size cannot exceed a threshold $L$. 
(3) \textbf{Number of connected components}: Maximize the number of connected components in the remaining network by deleting the fewest nodes possible, ensuring it meets or exceeds a threshold $N$. 
It discusses greedy algorithm design rules for these problems and proposes general algorithms for solving various CNPs, providing solutions for subsequent derivations and improvements of CNP models.
Alozie et al.~\cite{alozie2021efficient} focused on optimizing distance-based connectivity metrics by removing a subset of nodes. They aimed to minimize two specific metrics: (1) the number of node pairs connected by the longest path, and (2) the Harary index or efficiency of the graph. To achieve this, they proposed a general mixed-integer programming formulation and a separation algorithm based on breadth-first search tree generation, designed to solve the constraint inequalities.

Thulasiraman et al.~\cite{Thulasiraman} evaluated node value using cut vertices and edges for network diagnostics and security, employing link testing and node connection statistics.
In protein interaction networks, Boginski et al.~\cite{Boginski0007} aimed to identify key nodes by solving the Cardinally Constrained-CNDP (CC-CNDP), defined as minimizing the set of deleted nodes $R$ such that the largest connected component in the remaining subgraph $G(V \setminus A)$ of the network cannot exceed $L$, as follows:
\begin{equation}
\begin{aligned}
& \text{minimize} \quad \sum_{i \in V} v_i \\
\text{s.t.} \quad & 
 e_{ij} + v_i + v_j \geq 1, \quad \forall (i, j)\in E, \quad (1) \\
& e_{ij} + e_{jk} - e_{ki} \leq 1, \quad \forall (i, j, k)\in V, \quad (2) \\
& e_{ij} - e_{jk} + e_{ki} \leq 1, \quad \forall (i, j, k)\in V, \quad (3)\\
& -e_{ij} + e_{jk} + e_{ki} \leq 1, \quad \forall (i, j, k)\in V, \quad (4)\\
& \sum_{i, j \in V} e_{ij} \leq L, \quad (5)\\
& v_i \in \{0, 1\}, \quad \forall i \in V, 
 e_{ij} \in \{0, 1\}, \quad \forall i, j \in V, \quad (6)
\end{aligned}
\end{equation}
where $e_{ij}$ represents the edge between nodes $v_i$ and $v_j$, (1) ensures that there is an edge between nodes $v_i$ and $v_j$, meaning that both nodes are in the network, (2-4) constrain the connectivity of the node set $(i, j, k)$ to be minimal, and (5) constrains the size of the largest connected component in the network. This CNDP problem is transformed into an integer programming optimization problem to identify critical network disruptions and target proteins for drug neutralization. In self-organizing networks, Karygiannis et al.~\cite{Karygiannis1644271} defined critical nodes as breakpoints under failures or attacks. They used graph theory's vertex and edge cut principles to reveal network vulnerability through subgraph generation and involved solving optimization problems where their removal satisfies specific constraints.

\begin{table}[htbp]
 \centering
 \caption{Comparison of Control Optimization-Based Ranking Methods.}
\label{table06}
 \renewcommand{\arraystretch}{1.2}
 \setlength{\tabcolsep}{6pt}
 \begin{tabular}{m{4cm}m{4cm}m{7cm}}
 \toprule
 Title & Related Works & Advantages / Disadvantages \\
 \midrule
 Network Control & 
~\cite{li2006controlling}, 
~\cite{ghosh2022synchronized}, 
~\cite{d2023controlling}, 
~\cite{liu2011controllability}, 
~\cite{ding2017key}, 
~\cite{lu2019pinning}, 
~\cite{Zhu10106394}, 
~\cite{yu2012distributed}, 
~\cite{yu2013synchronization}, 
~\cite{ding2016optimizing}, 
~\cite{amani2017finding}, 
~\cite{amani2018controllability}, 
~\cite{liu2018optimizing}, 
~\cite{wang2012optimizing}, 
~\cite{bof2016role}, 
~\cite{zhou2018node}, 
~\cite{hui2021node}, 
~\cite{bomela2023finding}, 
~\cite{jiang2024pinning}, 
~\cite{sun2025optimal}, 
~\cite{zhang2017effect}, 
~\cite{mo2019effects}, 
~\cite{jiang2023searching}, 
~\cite{cao2024synchronization}, 
~\cite{zhang2022necessary}, 
~\cite{gao2025effects}
 & 
 $+$ Beneficial for dynamic networks and control applications, quantifies node controllability. 
 
 $-$ Requires accurate network topology information, relies on control theory. \\
 \midrule

Maximization of Connected  

Component Optimization  & \cite{aringhieri2016general},
~\cite{alozie2021efficient},
~\cite{Thulasiraman}, 
~\cite{Boginski0007},
~\cite{Karygiannis1644271}, 
~\cite{arulselvan2011cardinality}, 
~\cite{li2011finding},
~\cite{ventresca2015experimental}, 
~\cite{veremyev2014exact}, 
~\cite{ren2020identifying}, 
~\cite{wang2010community}, 
~\cite{lam2013identifying},
~\cite{ventresca2014fast},
~\cite{tang2020discrete} 
 & 
 $+$ Increases network fragmentation, applicable to security and IDS applications.
 
 $-$ Excessively breaks the network, affecting normal operations. \\
 \midrule
Minimization of Pair 

Connectivity Optimization & 
~\cite{di2012branch}, 
~\cite{ARULSELVAN20092193},
~\cite{shen2012adaptive},
~\cite{ventresca2014derandomized},
~\cite{dinh2015assessing},
~\cite{sarker2019critical}, 
~\cite{purevsuren2016heuristic}, 
~\cite{addis2016hybrid}, 
~\cite{chen2020critical}, 
~\cite{ventresca2012global},  
~\cite{shen2012discovery}, 
~\cite{ventresca2014region},  
~\cite{yin2023mixed23}, 
~\cite{zhang2024evolutionary}, 
~\cite{fortz2024min}, 
~\cite{jiang2025identifying},  
~\cite{kouam2025novel},  
~\cite{zhou2024finding}
 & 
 $+$ Effective in suppressing spreading and network attacks. 
 
 $-$ High computational demand in large-scale network. \\
 \midrule

 Minimum Spanning Tree & 
~\cite{chen2004evaluation}, 
~\cite{di2011complexity}, 
~\cite{hermelin2016parameterized}, 
~\cite{addis2013identifying}, 
~\cite{aringhieri2016local},
~\cite{wang2017evaluation}
 & 
 $+$ Simple and computationally efficient. 
 
 $-$ cannot fully capture network complexity and dynamics. \\
 \bottomrule
 \end{tabular}
\end{table}

Furthermore, Arulselvan et al.~\cite{arulselvan2011cardinality} proposed a risk management-based method to detect a set of nodes that, when deleted, cause the connectivity index of the resulting subgraph to fall below a threshold. This problem is defined as the cardinality-constrained critical node problem (CC-CNP) and modeled as an integer linear programming problem, which is solved using a heuristic genetic algorithm.
In social networks, there exists a set of $k$ mediator nodes aimed at maximizing the probability of information activation from a source node to a target node. Li et al.~\cite{li2011finding} defined this problem as the k-Mediators problem. Based on the independent cascade propagation model, they optimized information propagation by considering path probability effects and structural connectivity, using the product of edge weights to more accurately simulate the information spread process. They proposed a three-step greedy method to iteratively construct the best propagation tree (BPT), connecting all source and target nodes and optimizing the propagation paths to maximize activation probabilities.
Ventresca et al.~\cite{ventresca2015experimental} developed a multi-objective critical node detection algorithm. It removes nodes to maximize the connected components in the remaining network while minimizing the size variance among these components.
Veremyev et al.~\cite{veremyev2014exact} introduced two critical node deletion problems: (1) deleting $K$ nodes to minimize the remaining connected node pairs; (2) deleting the fewest nodes to ensure the largest component has no more than $L$ nodes. They reduced the solving time through constraints, making it possible to tackle large-scale network optimization issues.
Ren et al.~\cite{ren2020identifying} proposed a reverse greedy strategy to identify nodes crucial for maintaining connectivity. Starting from an empty network, it adds nodes with the least impact on the largest component, ranking by addition order, with the last added node deemed most critical.

To address the high computational costs of greedy algorithms on large networks.
Wang et al.~\cite{wang2010community} developed a community-based greedy algorithm using node edge weights for propagation speed, incorporating information diffusion for community detection and dynamic programming to identify influential nodes. Lam et al.~\cite{lam2013identifying} used simulations in the NetLogo multi-agent environment to identify critical infrastructure components, assessing node failure impacts via attack scenarios and the giant component metric. Ventresca et al.~\cite{ventresca2014fast} proposed a depth-first search method with $O(|V| + |E|)$ complexity, integrated into greedy algorithms for quick key node identification in large networks, enabling efficient CNDP objective function solving.
Tang et al.~\cite{tang2020discrete} introduced a Discrete Shuffling Frog Leaping Algorithm based on virtual frog population topology, featuring novel encoding, evolution rules, and a local degree-based replacement strategy for efficiency.

\subsection{Minimization of Pair Connectivity}\label{sec41}
To minimize pairwise connectivity by removing a fixed number of nodes, Di Summa et al.~\cite{di2012branch} proposed an integer linear programming (ILP) model with non-polynomial constraints, as follows:
\begin{equation}
\begin{aligned}
& \text{maximize} \quad \sum_{i, j\in V, i \ne j} e_{ij} \\
\text{s.t.} \quad & \sum_{i \in V} v_{i} \le K, \quad (1) \\
 \quad & v_{i} + v_{j} \ge e_{ij}, i, j \in E, \quad (2) \\
 \quad & e_{ij} + e_{jk} - e_{ik} \ge 0, i, j, k \in K, i < j<k, \quad (3) \\
 \quad & e_{ij} - e_{jk} + e_{ik} \ge 0, i, j, k\in V, i < j < k, \quad (4) \\
 \quad & -e_{ij} + e_{jk} + e_{ik} \ge 0, i<j<k, \quad (5) \\
 \quad & v_{i} \in \{0, 1\}, e_{ij}~\in\{0, 1\}, i, j \in V, i<j, \quad (6)
\end{aligned}
\end{equation}
where constraint (1) limits the number of deletions to $K$, while (2) ensures that removing an edge implies removing both its endpoints. Constraints (3)–(5) enforce connectivity dependencies, ensuring that if $\{i, j\}$ is disconnected in the remaining graph, then the node pairs $(i, k)$ and $(k, j)$ cannot be connected.

In small-scale networks, heuristic algorithms are commonly used to identify critical nodes that impact connectivity. Arulselvan et al.~\cite{ARULSELVAN20092193} employed a combinatorial heuristic integrating integer programming to efficiently minimize pairwise connectivity by selectively removing nodes. Similarly, Shen et al.~\cite{shen2012adaptive} introduced an adaptive algorithm that identifies critical nodes without requiring full recalculations after topological changes, enhancing efficiency and response speed.
Ventresca et al.~\cite{ventresca2014derandomized} developed an approximation algorithm for undirected, unweighted graphs, combining integer linear programming with randomized rounding. By handling vertex and triangle inequality constraints separately, their method effectively reduces subgraph connectivity while capping the number of deleted nodes at $k$, with solution quality bounded by a constant factor.
Dinh et al.~\cite{dinh2015assessing} assessed network vulnerability under dynamic conditions using expected pairwise connectivity (EPC) as a metric, transforming the problem into a stochastic optimization framework based on probabilistic graphs.
Sarker et al.~\cite{sarker2019critical} proposed an enumeration algorithm that accounts for connection costs and node weights, achieving high time efficiency and controlled complexity. Their approach, applied to river drainage networks, confirms power-law fragmentation behavior following the removal of critical nodes.

To address the complexity and accuracy balance in identifying critical nodes in large-scale networks, Pullan et al.~\cite{purevsuren2016heuristic} proposed a Greedy Randomized Adaptive Search Procedure (GRASP) with Path Relinking. The algorithm introduces an evolutionary path relinking mechanism during the path search process. It was shown that the GRASP with Path Relinking is more efficient than algorithms like neighborhood search and simulated annealing in solving the Critical Node Deletion Problem. 
Addis et al.~\cite{addis2016hybrid} proposed a "hybrid" heuristic algorithm based on the basic greedy algorithm. By combining two greedy rules, "node addition and deletion, " the algorithm addresses the CNDP problem of removing K nodes to minimize the residual network connectivity. The algorithm alternates between node addition and deletion operations in the feasible solution space, effectively avoiding local optimal solutions.
Chen et al.~\cite{chen2020critical} focused on the critical node fragmentation problem in weighted networks. They proposed a non-convex mixed-integer quadratic programming (MIQP) model for undirected weighted networks. A greedy algorithm is employed to approximate the optimal solution, and the algorithm iteratively selects K critical nodes (Critical Node Set, CNS) that cause the most significant decrease in network pairwise connectivity, identifying the critical node combination that maximizes network performance.

To mitigate the high computational complexity of the critical node deletion problem, 
Ventresca et al.~\cite{ventresca2012global} employed a population-based incremental learning algorithm combined with combinatorial simulated annealing to detect critical nodes, leveraging depth-first search to minimize pairwise connectivity. 
For dense networks where pairwise connectivity computations remain $O(n^{2})$, Shen et al.~\cite{shen2012discovery} developed a hybrid iterative linear programming rounding (HILPR) algorithm, which transforms the problem into ILP, improving solving efficiency through local search and constraint pruning. Ventresca et al.~\cite{ventresca2014region} proposed a region-growing dual-criterion method using LP relaxation to iteratively expand a fixed-radius ball, providing a logarithmic approximation for optimal node removal.
Addressing weighted social networks, Yin et al.~\cite{yin2023mixed23} introduced the Mixed Strength Decomposition (MSD) method, dynamically setting thresholds for node removal based on connection weights, improving the precision of critical node identification.
These advancements enhance computational efficiency in solving CNDP, enabling more effective identification of critical nodes in large-scale networks.

Recent studies have focused on optimizing critical node detection and network robustness under adversarial conditions. Zhang et al.~\cite{zhang2024evolutionary} formulated the Critical Node Detection problem in interdependent networks as a Bi-objective Optimization Problem (BICND) and designed a method integrating single-layer and multi-layer network transfer. This approach reduced algorithmic complexity while enhancing accuracy in identifying critical nodes. Fortz et al.~\cite{fortz2024min} proposed a compact integer programming formulation for the node attack optimization problem using pseudo-components within a two-level model, addressing performance limitations of prior iterative methods.
Jiang et al.~\cite{jiang2025identifying} introduced the EBC method, leveraging loop structures to assess network robustness against targeted attacks and precisely identify edges critical to resilience. Kouam et al.~\cite{kouam2025novel} simulated adversarial behavior to evaluate node influence, modeling attack progression toward the target. Zhou et al.~\cite{zhou2024finding} examined the NP-hard nature of the protection node problem and optimized key node selection via an influence redundancy mechanism. By minimizing the adjacency matrix’s spectral radius after key node removal, they demonstrated submodular properties in the objective function and achieved an approximation ratio of $O(N\log N)$.
These advances enhance network resilience by refining critical node and edge, offering efficient solutions for targeted attacks.
The multi-area power system can be described using power grid models and associated control models. In contrast, the networked $T-S$ fuzzy system is modeled using $T-S$ fuzzy models that consist of multiple fuzzy rules. The research on "Critical-metrics-based attack strategy and resilient $H_\infty$ state estimator design for multi-area power systems" presents a critical-metrics-based attack strategy. From the attacker's perspective, this strategy targets the critical information packets in the key areas of the power system, thereby amplifying the impact of the attack on system performance. Correspondingly, from the defender's perspective, a resilient state estimator is designed to alleviate the adverse effects of potential critical-metrics-based attacks.

\subsection{Minimum Spanning Tree}\label{sec42}
Chen et al.~\cite{chen2004evaluation} evaluated the relative importance of two sets of nodes by comparing the number of spanning trees in the remaining network after node deletion. The fewer the number of spanning trees in the corresponding network after removing a set of nodes and their associated edges, the more important that set of nodes is considered to be. This method is simple to compute and provides a more accurate reflection of node importance.
When the network size is large, the algorithm complexity of solving the problem increases. To improve the solving efficiency, the objective function is decomposed into multiple subproblems.
Di Summa et al.~\cite{di2011complexity} studied the minimization of the number of connected node pairs in the remaining graph after node deletion, designing the network as a tree structure to extend the solution to the CNP subproblem.
Hermelin et al.~\cite{hermelin2016parameterized} discussed the Critical Node Cut (CNC) problem: given a graph $G$ with $n$ vertices, determine $k$ nodes, and after their removal, the remaining graph has at most $x$ connected node pairs. Using a parameterized method, they proved that the problem can be solved in polynomial time by preprocessing it into a subproblem with lower complexity. 

Regarding the linear framework of the network traffic and connectivity constraint problem, Addis et al.~\cite{addis2013identifying} focused on the combinatorial problem of node deletion. They provided a dynamic programming recursive method for solving the tree decomposition problem based on graphs, which can solve the CNP problem in polynomial time on general and special graphs (split graphs, bipartite graphs, and the complements of bipartite graphs).
Aringhieri et al.~\cite{aringhieri2016local} proposed two metaheuristics based on iterative local search and variable neighborhood search frameworks to achieve maximum fragmentation of graphs. They designed two computationally efficient neighborhoods, evaluated different exploration strategies, and found that the improved strategy produced a large number of new best solutions. Using betweenness centrality, they were able to reduce the runtime without compromising the solution quality. Wang et al.~\cite{wang2017evaluation} combined the advantages of the minimum spanning tree (MST) structure and the minimum connected dominating set (MCDS) properties, introducing edge weight normalization to calculate the connection strength between nodes. They defined the minimum connected dominating set efficiency (SEW) and excluded non-critical endpoints, further distinguishing node importance.

\section{Machine Learning-based Ranking Methods}
Machine learning-based ranking methods have transformed the identification of influential nodes in complex networks by leveraging advanced techniques such as information entropy, clustering coefficients, graph neural networks, and reinforcement learning to capture intricate structural patterns and relationships. These methods find applications in diverse domains, including social network analysis, information diffusion, network security, and infrastructure management as details in Table.~\ref{table07}. Compared to traditional approaches, machine learning enables more accurate and nuanced assessment of node importance, while also being capable of handling large-scale networks and capturing nonlinear dependencies that conventional methods may overlook. However, these approaches introduce new challenges, such as the requirement for extensive training data, dependence on model design and generalization capability, and limitations in computational resources.

\subsection{Information Entropy}\label{sec50}

Entropy-based measures quantify the uncertainty or diversity in information propagation and have been widely applied to assess node centrality. Nikolaev et al.~\cite{nikolaev2015efficient} introduced entropy centrality, which evaluates a node's potential for dissemination via discrete Markov processes. Specifically, a node’s entropy centrality is defined as
\begin{equation}
 H^{t}_{i} = -\sum_{j=1}^{N}(p_{ij}^{(t)} + p_{ij'}^{(t)}) \log_2(p_{ij}^{(t)} + p_{ij'}^{(t)}).
\end{equation}
The term $p_{ij}^{(t)} + p_{ij'}^{(t)}$ represents the probability that an object originates from node $v_i$ and that during $t$ time periods, the node itself is found to possess node $v_j$. Higher entropy indicates a more uniform spread, implying greater unpredictability in the information’s destination. 
Building on this foundation, Zareie et al.~\cite{zareie2019influential} incorporated Shannon entropy and Jensen–Shannon divergence to develop the Diversity Strength Ranking (DSR) and Extended DSR (EDSR) algorithms, which account for both neighbor dispersion and influence range.

Other studies have leveraged entropy in various contexts. Nitt et al.~\cite{nitt2016using} proposed a degree centrality mapping entropy (ME) to identify critical nodes through local interactions. 
Similarly, Fu et al.~\cite{fu2015identifying} defined global node information entropy based on the $k$-shell index.
To address the issue in the VoteRank algorithm where all nodes are roughly assigned the same voting capacity and attenuation factor during voting, Guo et al.~\cite{guo2020influential} addressed limitations in VoteRank by introducing the EnRenew algorithm, which integrates connection-based entropy and an attenuation factor to update nodes’ propagation capabilities.
In response to the susceptibility of road networks to cascading failures caused by traffic accidents or abnormal events, Xu et al.~\cite{xu2018discovery} applied entropy concepts in the Origin-Destination Entropy and Flow (ODEF) and Crossroad Ranking (CRRank) to evaluate node importance using vehicle trajectory data. 
Tulu et al.~\cite{tulu2018identifying} further developed a Community-based Mediator (CbM) method that combines internal and external density entropy with node degree to identify critical mediators in large-scale networks.

Network entropy is typically used to characterize the amount of information encoded in a network structure. Based on the assumption that removing more important nodes may lead to greater structural changes, Ai et al.~\cite{ai2017node} defined node importance as the change in network entropy before and after its removal. Wu et al.~\cite{wu2023identify} proposed $k$-order entropy combined with betweenness centrality to capture global and local information through different order neighbor dependencies.
 Based on K-shell, the internal and external effects of neighbor subnet structure are defined as information entropy and cross-entropy~\cite{tong2023vital}, respectively, to identify the information disseminator. Li et al.~\cite{li2019key} further extended this approach by incorporating mutual information to derive a centrality measure (EMI) that considers both direct and second-order neighbor effects, effectively quantifying redundant and complementary information flows. These entropy-based approaches offer a rigorous framework to evaluate node importance by measuring the uncertainty and diversity inherent.

\subsection{Clustering Coefficient}\label{sec51}

The clustering coefficient is a crucial factor in determining the local influence of nodes. Chen et al.~\cite{RN35} combined the number of node neighbors, the influence of neighbor nodes, and the clustering coefficient to propose the ClusterRank method for local node importance ranking, specifically defined as:
\begin{equation}
 s_{i} = c(v_{i})\sum_{j\in \Gamma(v_i)} (k_{j}^{out} +1) .
\end{equation}
The local clustering coefficient $c(v_{i})$ of node $v_i$ is defined as: $c(v_i)=R(v_i)/k(v_i)(k(v_i)-1)$, where $R(v_i)$ represents the number of triangles formed by node $v_i$, and $\Gamma_{i}$ denotes the neighbors of node $v_i$. 
The term "+1" accounts for the degree contribution of the node itself to its neighboring nodes. 
When ranking nodes based on local centrality, traditional methods overlook the structural connections among neighbors. To address this, Gao et al.~\cite{gao2014ranking} proposed Local Structural Centrality (LSC), which integrates both the number of a node’s neighbors and their topological interconnectivity. By incorporating the local clustering coefficient, LSC quantifies the extent to which a node’s neighbors are interconnected, capturing structural nuances often ignored in purely degree-based metric. 
The local clustering coefficient reflects the density of connections among a node’s immediate neighbors, offering insights into its influence within the network. 
Nodes with similar local centrality values but higher clustering coefficients tend to exhibit stronger propagation capabilities, as their tightly connected neighborhoods facilitate more effective information diffusion.
In addition, the traditional K-truss method, which decomposes networks based on edge clustering, may be affected by “false cores” when clustering characteristics are not accurately captured. Yang et al.~\cite{yang2018identifying} proposed an improved K-truss approach that incorporates both edge diffusivity and edge clustering coefficients, thus providing a balanced metric for identifying influential nodes. Complementarily, Zareie et al.~\cite{zareie2020finding} developed the ECRM method, which posits that nodes with lower correlation to their neighbors are more likely to be influential, as they can more effectively disseminate information to the network periphery.

To overcome the limitations of undirected graphs in representing conditional independence, Dablander et al.~\cite{dablander2019node} developed a causal framework based on directed acyclic graphs (DAGs). Assuming the Markov condition holds, they introduced the concept of "faithfulness" to ensure that the DAG accurately captures variable independencies. They further quantified causal impacts using Average Causal Effect (ACE) and Kullback-Leibler (KL) divergence. Their findings indicate that eigenvector centrality outperforms traditional metrics in predicting the causal influence of nodes within such networks. Liu et al.~\cite{liu2022identifying11} integrated latent group information in power systems by developing the Group-Driven Framework for Identifying Critical Nodes (GDF-ICN) algorithm. This framework combines the structural and operational states of power systems with node group affiliations, iteratively optimizing group partitions around critical nodes. A fuzzy closeness metric is incorporated to enhance the description of group structure, while leveraging clustering effects to bolster the robustness of critical node identification. Wang et al.~\cite{wang2022influential22} proposed an influential node evaluation algorithm grounded in information entropy theory. By defining metrics based on the number of triangles and edge weights, they introduced the concept of edge entropy weights—capturing differences in edge weights and their influence on neighboring nodes. This approach integrates k-core metrics to characterize local node influence, with higher entropy values indicating greater complexity and overall influence within the network. Zhang et al.~\cite{zhang2013identifying} presented a novel method focused on maximizing the range of influence. Their approach constructs an information transmission probability matrix, which quantifies the probability of information transfer between arbitrary node pairs.

\subsection{Machine Learning}
\label{sec52}
Machine learning offers a powerful framework for identifying influential nodes by reframing the problem as a regression task. These methods leverage supervised learning algorithms trained on node labels derived from epidemic model simulations and traditional centrality measures, capturing complex relationships between network structure and propagation potential.
Zhao et al.~\cite{zhao2020machine} employed information infection vectors to encode topological characteristics and infection dynamics across diverse propagation scenarios. Using node labels obtained from SIR model simulations, they trained classifiers and regressors—including Naïve Bayes, decision trees, random forests, SVM, KNN, logistic regression, and MLP—to predict node importance. Yang et al.~\cite{yang2021identifying36} introduced a network-aware local centrality index (NLC) that integrates topology with node embeddings. Applying DeepWalk to project network structures into a low-dimensional space, they incorporated third-order neighborhood information to refine influence estimation. Rezaei et al.~\cite{rezaei2023machine356} further advanced this paradigm by engineering collective node features—spanning connectivity, degree, and coreness measures—and training an SVR model with an RBF kernel, demonstrating the efficacy of tailored feature representations in predicting node influence.
However, existing models face limitations in feature representation and scalability. The development of deep learning and graph neural network approaches provides a powerful new framework for analyzing node attributes and quantifying influence.

\textbf{Graph Conventional Network:} Zhao et al.~\cite{ZHAO202018} pioneered the deep learning framework InfGCN based on graph convolutional networks (GCN). InfGCN constructs node structural features—including degree, closeness, betweenness, and clustering coefficient—and builds a fixed neighbor network via breadth-first search (BFS). By training with actual infection rate vectors from SIR model simulations, InfGCN minimizes the error between predicted and actual infection rates to accurately estimate node influence. 
Similarly, Kumar et al.~\cite{kumar2022influence23} adopted the Stru2vec node embedding algorithm~\cite{ribeiro2017struc2vec} along with a message-passing neural network (MPNN) to process node embeddings and SIR-based influence vectors, identifying the Top-$k$ influential nodes.
Zhang et al.~\cite{zhang2022new57} combined convolutional neural networks (CNN) with GNNs in their CGNN algorithm, which simplifies feature matrices and focuses on first- and second-order neighbors to label nodes via the SIR model and optimize a loss function for precise identification of critical nodes. Liu et al.
~\cite{liu2022learning} developed a ranking method based on self-supervised learning and graph convolutional models, leveraging both node features and network structural information.

\textbf{Graph Embeddings:}
Wei et al.~\cite{wei2024enimnr} focused on reducing computational complexity. They selected candidate nodes based on shell connections and topological features, then used deep learning to generate low-dimensional vectors for calculating propagation dependency. Keikha et al.~\cite{keikha2020influence} proposed DeepIM, which extracts global and local structural features via CARE, constructs custom paths, and employs Word2vec to learn node feature vectors for measuring user relevance. Bouyer et al.~\cite{bouyer2024maximizing} presented ETIM, combining shell decomposition, graph embedding, and local structure features to decrease computation. Wu et al.~\cite{YihangandHu2024} constructed node features using rule-equivalent similarity and Graph Convolutional Networks. Ahmad et al.~\cite{ahmad2024learning} proposed frameworks (LCNN) that merge CNNs with local node representations and multi-scale metrics. Rashid et al.~\cite{rashid2024olapgn} introduced the GN model, integrating deep learning and probabilistic properties via Graph Convolutional Networks to detect overlapping communities and key nodes. Xiong et al.~\cite{xiong2024vital} proposed the AGNN algorithm, fusing autoencoders with GNNs to generate topological feature embeddings through GCN and optimizing the ranking prediction model using listMLE. Yu et al.~\cite{yu2020identifying} developed methods RCNN that integrate GCNs and CNN-based adjacency features for effective critical node detection.

\textbf{Graph Attention Network:} Most existing node importance assessments in knowledge graphs fail to fully utilize available information and cannot capture node relationships and attribute features. To address this, 
Park et al.~\cite{park2019estimating} GENI uses predicate-aware attention mechanisms and flexible centrality adjustments to aggregate node and neighbor information across multiple GNN layers. It updates node embeddings via an aggregation function with a nonlinear transformation and maps these to importance scores using a scoring neural network. GENI also incorporates node in-degree centrality to adjust importance estimates.
Define a GNN with $L$ layers of networks, where the $l$-th aggregates information from the $l-1$ th layer as follows, 
\begin{equation}
\begin{aligned}
 & \overrightarrow{h}_{N(i)}^{l} \gets \text{TransForm}^{l} \left ( \text{Aggregate} \left ( \left \{\left (\overrightarrow{h}_{j}^{l-1}, w_{i, j}^{l} \right ) | j\in N(i) \right \} \right ) \right ), \\
 & \overrightarrow{h}_{i}^{l} \gets \text{Combine} \left( \overrightarrow{h}_{i}^{l-1}, \overrightarrow{h}_{N(i)}^{l} \right), 
\end{aligned} 
\end{equation} 
where $\overrightarrow{h}_{i}^{l-1}$ is the $v_i$ represents the information aggregation of the $(l-1)$ th layer, each neighbor node $N(i)$ of node $v_i$ denotes the set of its neighbors (each initialized with a weight $w_{i, j}^{l}$). The aggregation function and nonlinear transformation function share parameters across all nodes. At each layer, node scores are updated through a weighted combination of their own value and neighbors' values, with attention mechanisms dynamically adjusting connection influence. Initial node scores come from feature embeddings processed through a neural scoring network. GENI also incorporates a centralization score based on node in-degree, refined through scaling and offset adjustments. Finally, a nonlinear activation function integrates this score with learned representations.
Furthermore, Park et al.~\cite{park2020multiimport} designed MultiImport, an end-to-end latent variable model to infer node importance from multiple sparse and possibly overlapping input signals. MultiImport uses edge-aware attention aggregation and node centralization adjustment within a GNN framework to learn robust node representations. It models the knowledge graph structure and uses clustering to group nodes with similar functional signals.  Munikoti et al.~\cite{munikoti2022scalable} introduced ILGR, a scalable GNN-based framework that approximates node and link importance through attention-driven embedding. ILGR encodes local structure by aggregating neighborhood information, with an attention mechanism assigning adaptive weights based on structural roles. Ge et al.~\cite{ge2024node} proposed MFA-NIE, integrating structural, relational, and attribute features with an enhanced attention mechanism for feature extraction, and adjusts node importance using TOPSIS centrality. Chen et al.~\cite{chen2024social} developed a hierarchical graph-enhanced variational autoencoder that reconstructs neighborhood representations to learn social influence patterns, using hierarchical graph attention networks to capture influential neighborhoods and mitigate data sparsity.
Liu et al.~\cite{liu2022nie23} introduced NIE-GAT for inter-domain routing networks, mapping static node features to dynamic cascading failure impacts through graph attention networks. Additionally, 
Kou et al.~\cite{kou2023identify23} refined GCN-based approaches by framing node influence prediction as a regression task and using a multi-head attention mechanism to integrate neighbor importance, with the IC model providing simulation-based ground truth for optimization.

\textbf{Graph Contrast Learning:}
Recent advancements in contrastive learning have improved node importance estimation by integrating multi-view representations and attention mechanisms. Liu et al.~\cite{liu2023node} addressed the limitations of single-graph models by introducing the MCRL framework, which employs dual graph encoders and contrastive learning to derive multi-perspective node representations. This approach enhances the estimation of node importance by comparing embeddings generated through graph convolutional networks (GCN) and graph attention mechanisms (GAT). By incorporating a positive-negative sampling strategy, MCRL identifies nodes that maintain both high similarity and distinctiveness across multiple structural views—key indicators of importance.
Building on this concept, Zhang et al.~\cite{zhang2024label} developed the Label-Contrastive Pretraining (LICAP) model, which employs hierarchical sampling to prioritize top-ranked nodes. LICAP utilizes a Predicate-Aware Graph Attention Network (PreGAT) to refine node embeddings, ensuring robust differentiation between influential and peripheral nodes. Meanwhile, Shu et al.~\cite{shu2025node} proposed AGCL, a contrastive learning framework for heterogeneous networks that leverages attention mechanisms to quantify both local and global node importance. By extracting structural features through contrastive learning and computing importance weights via attention mechanisms, AGCL offers a comprehensive approach to evaluating node influence in complex networks.

\textbf{Graph Neural Networks:}
Huang et al.~\cite{huang2022estimating} introduced HIVEN, a GNN-based framework designed for heterogeneous information networks (HINs). By integrating a heterogeneous information aggregator and a meta-path-based mechanism, HIVEN effectively captures the structural and relational intricacies of HINs, improving both accuracy and efficiency in node ranking.
Building on structural knowledge integration, Chen et al.~\cite{chen2024deep} developed SKES (Deep Structural Knowledge Exploitation and Synergy), which incorporates node centrality and similarity metrics to learn high-dimensional feature distributions. SKES refines importance predictions using optimal transport theory, enabling precise characterization of node significance within heterogeneous graphs. Similarly, Lin et al.~\cite{lin2024node} leveraged Large Language Models (LLMs) in the LENIE framework to enhance semantic representations in knowledge graphs. By employing a clustering-based triplet sampling strategy and node-specific adaptive prompts, LENIE enriches node embeddings, thereby improving the initialization of node importance estimation models.
Higher-order network structures have also been incorporated into learning-based ranking methods. Zhao et al.~\cite{zhao2023novel} proposed HONNMA, a high-order neural network framework that employs motif attention to capture complex topological dependencies. This model encodes interactions through a weighted motif adjacency matrix and refines node embeddings using a skip-connected architecture, allowing for more robust node importance estimation. Addressing the Critical Node Detection Problem, Michos et al.~\cite{michos2024critical} applied Hopfield Neural Networks (HNN) to optimize network resilience. By reformulating as an energy minimization task, their approach identifies critical nodes by converging toward stable network states with minimal energy.

\subsection{Reinforcement Learning}\label{sec53}
Reinforcement learning influences key node selection by using rewards and penalties to guide the process, treating the impact of nodes—whether on overall network performance or individual propagation capability—as the value incentive. In this framework, the objective is to identify a set of important nodes by either maximizing or minimizing a value-based objective function.
To tackle high time complexity in critical node identification, Fan et al.~\cite{fan2020finding} proposed the FINDER framework. It minimizes the accumulated normalized connectivity (ANC) through a node deletion strategy. The ANC is defined as:
\begin{equation}
R(v_{1}, v_{2}, \dots, v_{N})=\frac{1}{N} \sum_{k=1}^{N} \frac{\sigma(\mathscr{G}~\setminus \{v_{1}, v_{2}, \dots, v_{k})\}}{\sigma(\mathscr{G})} c_{v_{k}}, 
\end{equation}
where $c(v_{k})$ is the normalized removal cost of node $v_{k}$, and $\sigma$ measures network connectivity. This optimization process is modeled as a Markov Decision Process (MDP), defined by the following components: state space $\mathcal{S}$ represents the residual graph $\mathscr{G}'$ after a sequence of node removals, action space $\mathcal{A}$ denotes the selection of a node $v\in V(\mathscr{G}')$ for removal at each step, and reward function $r:\mathcal{S}\times \mathcal{A}$ is the reduction in ANC resulting from removing node 
$v$, incentivizing actions that lead to rapid network fragmentation. A strategy for selecting the next node to delete, updated iteratively via reinforcement learning algorithms as policy $\pi(a|s)$. The iterative process involves evaluating candidate actions based on their expected rewards, selecting the node that maximizes the reward signal, and updating model parameters using a greedy or approximate policy improvement strategy. To further reduce the computational complexity of heuristic algorithm-based node deletion, Tan et al.~\cite{tan2024learning} proposed a feature-sensitive graph attention network. This method represents nodes by their features and, combined with an adversarial double-depth network, constructs a flexible end-to-end framework for generalizing critical node identification across different scales.

To address information diffusion features in node importance analysis, several studies have integrated reinforcement learning with graph-based methods. Jaques et al.~\cite{jaques2019social} used a multi-agent reinforcement learning framework with counterfactual reasoning and measures like KL divergence to assess the impact of alternative actions on other agents, identifying influential nodes. Chen et al.~\cite{chen2023identifying79} proposed DeepELE, combining graph embeddings with reinforcement learning in an SIS model. Each node is assigned a diffusion weight, and a state–action value function evaluates the network-wide infection impact after node removal, selecting key nodes by minimizing an objective function.
Li et al.~\cite{li2019disco} introduced DISCO, combining network embedding with deep reinforcement learning for various node-related problems. Chen et al.~\cite{chen2023touplegdd} developed ToupleGDD, an end-to-end deep reinforcement learning framework coupling three graph neural networks with a double deep Q-network for parameter learning, addressing influence maximization while overcoming limitations like narrow IM formulations and low scalability.
Ling et al.~\cite{ling2023deep} introduced DeepIM, generating latent representations of seed sets in a data-driven manner to capture diverse information diffusion patterns, and designing a novel objective function for optimal seed sets under flexible node centrality constraints. Li et al.~\cite{li2022piano} presented PIANO, merging deep reinforcement learning with network embedding to estimate node influence, providing a pre-trained model pool for direct application. Uthayasuriyan et al.~\cite{uthayasuriyan2024adaptive} introduced DERL, integrating differential evolution with DQN-based reinforcement learning, and Li et al.~\cite{li2024adarisk} proposed AdaRisk, a risk-adaptive deep reinforcement learning framework for detecting fragile nodes in uncertain graphs.
Xu et al.~\cite{xu2025influence} developed HEDRL-IM, reformulating influence maximization in hypergraphs as a network weight optimization task solved via deep Q-networks, integrating evolutionary algorithms and propagation simulation. Zhu et al.~\cite{zhu2025bigdn} proposed BiGDN, employing bidirectional neighborhood aggregation and integrating deep reinforcement learning with multi-head attention for robust node representations. Ahmad et al.~\cite{ahmad2025learning} developed CoreQ, using K-core decomposition and Q-learning to identify candidate seed nodes.

\begin{table}[htbp]
 \centering
 \caption{Comparison of Machine Learning-Based Ranking Methods.}
\label{table07}
 \begin{tabular}{m{4cm}m{4cm}m{7cm}}
 \toprule
 Title & Related Works & Advantages / Disadvantages\\
 \midrule Information Entropy 
 &\cite{nikolaev2015efficient}, 
~\cite{zareie2019influential}, 
~\cite{nitt2016using}, 
~\cite{fu2015identifying}, 
~\cite{guo2020influential}, 
~\cite{xu2018discovery}, 
~\cite{tulu2018identifying}, 
~\cite{ai2017node}, 
~\cite{wu2023identify}, 
~\cite{tong2023vital}, 
~\cite{li2019key}
 & $+$ Robustness to various network structures. 
 
 $-$ Dependent on accurate data, hard for large or dynamic networks. 
 \\
 \midrule Clustering Coefficient 
 &~\cite{RN35}, 
~\cite{gao2014ranking}, 
~\cite{yang2018identifying}, 
~\cite{zareie2020finding}, 
~\cite{dablander2019node}, 
~\cite{liu2022identifying11}, 
~\cite{wang2022influential22}, 
~\cite{zhang2013identifying} 
 & $+$ Improves identification of influential nodes. 
 
 $-$ Potential oversimplification of network dynamics, reliance on local information. 
 \\
 \midrule Graph Conventional Network
 &~\cite{ZHAO202018}, 
~\cite{kumar2022influence23}, 
~\cite{zhang2022new57}, 
~\cite{liu2022learning}
 
 & $+$ Straightforward implementation, robustness in various network types. 
 
 $-$ Limited ability to capture complex patterns, potential inefficiency with large-scale networks.
 \\
 \midrule Graph Embeddings
 &\cite{wei2024enimnr}, 
~\cite{keikha2020influence}, 
~\cite{bouyer2024maximizing}, 
 ~\cite{YihangandHu2024}, 
 ~\cite{ahmad2024learning}, 
 ~\cite{rashid2024olapgn}, 
~\cite{xiong2024vital}, 
~\cite{yu2020identifying}
 & $+$ Improves accuracy for influential nodes. 
 
 $-$ Reliance on quality of embeddings.
 \\
 \midrule Graph Attention Network
 &\cite{park2019estimating}, 
~\cite{park2020multiimport}, 
~\cite{munikoti2022scalable}
~\cite{ge2024node}, 
~\cite{chen2024social}, 
~\cite{liu2022nie23}, 
~\cite{kou2023identify23}

 & $+$ Heightened accuracy in ranking influential nodes, the ability to focus on important features.
 
 $-$ Dependency on hyperparameter tuning, potential overfitting in small datasets.
 \\
 \midrule Graph Contrast Learning
 & 
~\cite{liu2023node}, 
~\cite{zhang2024label},
~\cite{shu2025node}
 & $+$ Effective identification of key nodes by leveraging differences between subgraph representations. 
 
 $-$ High computational cost, potential overfitting. 
 \\
 \midrule Graph Neural Networks
 &\cite{huang2022estimating}, 
~\cite{chen2024deep}, 
~\cite{lin2024node}, 
~\cite{zhao2023novel}, 
~\cite{michos2024critical}
 & $+$ Superior ability to capture complex patterns, adaptability to diverse network structures. 
 
 $-$ Dependence on extensive training data and difficulties in model interpretability.
 \\
 \midrule Reinforcement Learning
 &\cite{fan2020finding}, 
~\cite{tan2024learning}, 
~\cite{jaques2019social}, 
~\cite{chen2023identifying79}, 
~\cite{li2019disco}, 
~\cite{chen2023touplegdd}, 
~\cite{ling2023deep}, 
~\cite{li2022piano}, 
~\cite{uthayasuriyan2024adaptive}, 
~\cite{li2024adarisk}, 
~\cite{xu2025influence}
~\cite{zhu2025bigdn}, 
~\cite{ahmad2025learning}

 & $+$ Dynamic adaptation to network changes, improves decision-making over time. 
 
 $-$ Complexity in setting reward structures, reliance on extensive exploration and training periods.
 \\
 \bottomrule
 \end{tabular}
\end{table}

\section{Comprehensive Index Based Ranking Methods}
Comprehensive index ranking methods evaluate node importance by integrating multiple network attributes and advanced analytical techniques. These methods combine various centrality measures, gravity models, and decision-making frameworks to overcome the limitations of single-indicator approaches and provide a more holistic assessment of node significance in complex networks. These indicators are widely applied across domains such as social network analysis, biological networks, transportation systems, and infrastructure management. The primary advantage of these methods is their ability to incorporate diverse structural and functional indicators, thereby offering a more nuanced characterization of node importance. However, they also introduce challenges, including increased computational complexity and careful selection and integration of relevant indicators.

\subsection{Gravity Formula}\label{sec60}
The law of universal gravitation has been innovatively applied in network science. By defining the gravitational constant multiplied by the product of the masses of two nodes divided by the square of the distance between the nodes, the centrality and statistical indicators of nodes are considered as "mass", and the shortest distance between nodes is used as the distance, thus constructing a new framework for measuring node importance. Ma et al.~\cite{RN14} combined the K-shell index of nodes and the shortest path between nodes to identify important nodes in the network through the gravitational calculation formula, defining the importance of node $v_i$ as:
\begin{equation}
\label{ksgm}
 GM(v_{i})= \sum_{j\in \Psi } ks(i)\cdot ks(j) / d_{ij}^{2}, 
\end{equation}
where $ks(i)$ and $ks(j)$, denote the $k$-shell values of nodes $v_i$ and $v_j$, $d_{ij}$ is the shortest path between them, and $\Psi$ is the set of neighbors within a radius $r$ of node $v_i$. In parallel, Maji et al.~\cite{maji2020influential} addressed limitations stemming from empirically set free parameters in improved $k$-shell decomposition algorithms by considering the influence of different levels of neighbors, thereby enhancing the adaptability, efficiency, and accuracy of the model.

In addition, Li et al.~\cite{RN16} proposed the Local Gravity Model (LGM), which extends the basic gravity model to include neighborhood information; nodes surrounded by neighbors with high $k$-shell indices are considered closer to the network’s center. 
Li et al.~\cite{RN24}
further modified the model by replacing the $k$-shell metric with the degree of neighboring nodes. Recognizing that a degree-based gravity model captures only local information. Liu et al.~\cite{liu2020gmm} introduced the eigenvector exponent of the adjacency matrix into the gravitational formula, proposing the Generalized Gravity Model (GMM) that incorporates both local and global network characteristics.
Yang et al.~\cite{RN13} utilized $k$-shell values to define an attraction coefficient between nodes, taking into account critical positional attributes of a node’s neighborhood.
Li et al.~\cite{RN20} measured local information using each node's local clustering coefficient and clustering degree, thereby defining a node’s spreading ability as its mass derived from both neighbor information and node degree. In a related approach, 
Fei et al.~\cite{fei2018identifying} assessed node importance by defining mutual attraction based on the inverse-square law, though this method may overlook more complex inter-node relationships.

A single centrality measure often fails to fully capture the importance of nodes in a network. To address this, researchers have introduced weighted models based on gravitational principles, incorporating multiple attributes to assess node influence. Wang et al.~\cite{RN28} enhanced the GM method by integrating $k$-shell value and degree centrality while further considering neighbor information to improve accuracy. 
Building on this approach, Li et al.~\cite{RN26} proposed the MCCM method, which defines node importance as a weighted combination of degree, $k$-shell value, and eigenvector centrality. To ensure comparability across different networks, these metrics are normalized, with a balancing parameter introduced to prevent the $k$-shell value from disproportionately influencing smaller networks. Similarly, Yan et al.~\cite{yan2020identifying} adopted a multi-attribute decision-making approach, constructing a decision matrix based on multiple centrality measures and using entropy weighting to determine a node’s overall importance. Beyond centrality weighting, researchers have also addressed the issue of influence overlap among high-centrality nodes. Wang et al.~\cite{wang2022identifying20} introduced the centripetal centrality index, which integrates node attributes and connectivity to reduce redundancy in influence propagation, albeit at a high computational cost. Yang et al.~\cite{yang2023aigcrank17} developed the AIGCrank algorithm, combining gravitational centrality with a recursive ranking strategy that integrates the h-index, domain core centrality, and clustering coefficient, effectively distinguishing key nodes while mitigating influence overlap. Other approaches focus on probability-based propagation and structural optimization. Zhu et al.~\cite{zhu2023identifying12} introduced the HVGC method, incorporating the h-index, node distance, and an exponential decay factor to accurately identify bridge nodes that connect different network regions, even if they exhibit low $k$-shell values. Additionally, Liu et al.~\cite{liu2023entropy12} developed the SEGM and EGM models, respectively, leveraging information entropy to capture the diffusion capabilities of nodes more effectively.

To mitigate the high computational cost of shortest path calculations, researchers have proposed alternative approaches that approximate or replace shortest path metrics with more efficient methods. Zhao et al.~\cite{zhao2022random21} employed random walks to estimate node distances, using degree centrality as a proxy for node influence. Extending this idea, Zhao et al.~\cite{zhao2023estimating3} introduced the NEGM model, which replaces shortest path lengths with Euclidean distances derived from Node2Vec embeddings. Similarly, Shang et al.~\cite{RN25} defined a probabilistic distance using Markov chain-based transition probabilities, capturing latent inter-node relationships without computing exact shortest paths.
Beyond shortest path approximations, alternative gravity-based models have been proposed. Curado et al.~\cite{curado2023novel} introduced a method leveraging return-based random walks and effective distance to capture both static and dynamic network topologies. Yang et al.~\cite{yang2023aogc} refined interaction range definitions by incorporating an adaptive truncation radius and considering all-channel paths instead of shortest path constraints. Chen et al.~\cite{chen2023identification} proposed the DCGM model, integrating degree and average neighbor degree to enhance influence estimation.
Further advancements address asymmetric interactions and multilayer structures. Meng et al.~\cite{meng2025improved} developed the Asymmetric Attraction Model (AAM), transforming adjacency matrices into asymmetric attraction matrices to capture directional dependencies. Xu et al.~\cite{xu2024cagm} introduced the CAGM model, incorporating influence probability, $k$-shell, and degree information for adaptive gravity-based estimation. Li et al.~\cite{li2023improved123} extended gravity-based models to multilayer networks, proposing the APAMGM algorithm for multiplex transportation networks. Lü et al.~\cite{lv2024improved} introduced PRGC, integrating multi-PageRank centrality with weighted shortest path distances. Chi et al.~\cite{chi2024measuring} developed a semi-global centrality measure inspired by stellar models, incorporating variable propagation speeds and the gravitational slingshot effect to quantify node efficiency in information diffusion.

\subsection{Multiple Metrics}\label{sec61}


To address challenges in identifying critical nodes arising from varied network structures and densities, researchers have developed composite centrality measures that integrate multiple metrics. For instance, 
Comin et al.~\cite{comin2011identifying} proposed a comprehensive centrality index that balances global and local characteristics by analyzing three different information propagation strategies. 
De Arruda et al.~\cite{de2014role} further explored dynamic propagation by proposing the Random Walk Accessibility (RWA) expansion indicator, based on nine centrality measures, to assess the ability of nodes to spread information in both spatial and non-spatial networks.
To integrate multiple centrality measures, Hu et al.~\cite{hu2015multi} adopted linear discriminant analysis to integrate eigenvector, betweenness, closeness, degree centrality, and mutual information, developing a multi-metric evaluation algorithm. 
Hu et al.~\cite{hu2015identifying} then proposed the NICCM method, constructing a contribution matrix that defines importance by combining efficiency with weighted contributions.

Several studies have integrated multiple centrality measures to enhance prediction accuracy. Bucur et al.~\cite{bucur2020top} combined statistical classifiers into composite metrics, demonstrating high predictive power in SIR epidemic models. Wei et al.~\cite{wei2022identifying24} developed a heterogeneous mean-field model incorporating betweenness, degree, H-index, and core degree, showing that betweenness-based immunization strategies are effective in BA networks, whereas degree-based strategies perform well in real networks. An et al.~\cite{an2024novel} introduced the DIRCI method, leveraging dynamic influence scope, network layer centrality, and community centrality for critical node identification in multilayer networks. Wu et al.~\cite{wu2024hunting} proposed radiation centrality, which models information dissipation using attenuation and scattering theories.
Other approaches combine local and global structural features. Cao et al.~\cite{cao2024dynamic} introduced INLGC, integrating local network constraint coefficients with global community structure. Kopsidas et al.~\cite{kopsidas2022identification} merged centrality measures from different transport networks to assess metro station criticality. Wang et al.~\cite{wang2024multi} proposed a multi-factor information matrix centrality algorithm, incorporating node influence, neighbor influence, and feedback-based mutual influence. Lei et al.~\cite{lei2024weighted} developed the Weighted Information Index (WII), which utilizes second-order neighbor information to construct an information distribution vector. Ullah et al.~\cite{ullah2024leveraging} introduced NPIC, integrating neighborhood and path information, while Zhang et al.~\cite{zhang2024critical} employed the Comprehensive Voting Ranking (CVR) algorithm, incorporating mutual information and k-center clustering to identify critical nodes in urban rail networks.
Several iterative and hybrid models have also been proposed. Lee et al.~\cite{lee2024identifying} developed the Balanced Iterative Influence (BII) algorithm, iteratively combining local structural information with global influence. Esfandiari et al.~\cite{esfandiari2025collaborative} introduced HNPR, which integrates $k$-shell and PageRank into a novel linear metric. Chen et al.~\cite{chen2025sfimco} proposed the SFIMCO method, distinguishing overlapping from non-overlapping nodes to quantify intra-community influence. Zhang et al.~\cite{zhang2025novel} introduced the SLCMF metric, integrating structural, social, and semantic factors while improving scalability via distributed subgraphs and shortest path approximations.
Dempster-Shafer evidence theory has also been applied to node importance estimation. Mo et al.~\cite{mo2015evidential} developed Comprehensive Evidence Centrality (CEC), integrating degree, betweenness, and closeness centrality. Xu et al.~\cite{xu2025lgp} introduced the Local-Global-Position (LGP-DS) algorithm, which combines global, local, and positional attributes, leveraging information entropy to evaluate and aggregate their contributions to node importance.

To address the challenge of distinguishing nodes within the same $k$-shell layer, Sheikhahmadi et al.~\cite{sheikhahmadi2017identification12} introduced the MCDE hybrid index, integrating node shell, degree, and entropy information to enhance accuracy and efficiency.
Wang et al.~\cite{RN51, wang2017ranking}, Sheikhahmadi et al.~\cite{sheikhahmadi2017identification}, and Yang et al.~\cite{yang2018ranking} developed iterative removal-based indices that combine node position and factor indices to achieve efficient $O(n)$ complexity. Namtirtha et al.~\cite{namtirtha2021best} introduced the NGSC indicator, a weighted sum of $k$-shell values and neighbor degrees, adaptable to both dense and sparse networks.
Hybrid approaches integrating local and global structure have also been explored. Ullah et al.~\cite{ullah2021identification} developed the Global Structure Model (GSM) to capture global node influence, while the Local and Global Centrality (LGC) metric~\cite{ullah2021identifying12} combines $k$-shell entropy with neighbor degree sums. 
Hu et al.~\cite{hu2022exploring209} proposed a neighbor similarity coefficient, integrating $k$-shell hierarchy and node degrees for improved influence evaluation.
Wang et al.~\cite{wang2022influential} developed the ALSI method, incorporating both inherent and neighbor-driven influence. Mukhtar et al.~\cite{mukhtar2023integrating} introduced H-GSM, combining degree centrality and $k$-shell centrality to overcome single-metric limitations. Yang et al.~\cite{yang2023identifying23} refined influence ranking using a structural iteration factor method, distinguishing nodes through iteration steps to mitigate propagation limitations. Qiu et al.~\cite{qiu2024key} proposed a hierarchical composite algorithm, integrating constraint coefficients from the salton index with improved $k$-shell for local-global ranking.

\subsection{Topsis and Entropy-Weight}\label{sec62}
The Technique for Order Preference by Similarity to Ideal Solution (TOPSIS) method integrates multiple centrality metrics, offering a comprehensive and objective approach to evaluate node importance. Du et al.~\cite{du2014new} proposed a new method for assessing node importance in complex networks using the TOPSIS method. A decision matrix $D=(x_{mn})$ is constructed using the multiple attributes of nodes in the network, and the decision matrix is normalized as$ r_{ij} = x_{ij} / \sum_{j=1}^{m}x_{ij}^{2}, \quad i=1, \dots, m, j=1, \dots, n$ .
Then, multiply each column of this normalized matrix by its corresponding weight to obtain the weighted decision matrix $A=(v_{mn}): v_{ij}=w_{ij}~\times r_{ij}$
where $w_{j}$ is the weight of the $v_j$ column. The TOPSIS method selects the alternative that is simultaneously closest to the positive ideal solution and furthest from the negative ideal solution. 
Define the positive ideal solution as $A^{+}$ and the negative ideal solution as $A^{-}$, defined as follows:
\begin{equation}
\begin{aligned}
 A^{+} = \left \{v_{1}^{+}, v_{2}^{+}, \dots, v_{n}^{+} \right \} = \left \{\left( \max_{i}v_{ij} | j\in K_{b} \right ) \left( \min_{i}v_{ij} | j\in K_{c} \right ) \right\}, \\
 A^{-} = \left\{v_{1}^{-}, v_{2}^{-}, \dots, v_{n}^{-} \right\}
 = \left\{\left( \min_{i}v_{ij} | j\in K_{b} \right ) \left( \max_{i}v_{ij} | j\in K_{c} \right ) \right\}, 
\end{aligned}
\end{equation}
where $K_{b}$ and $K_{c}$ are benefit and cost sets. The Euclidean distances to these ideals, $S^{+}_{i}$ and $S^{-}_{i}$ were calculated, and the relative closeness $C_{i} = \frac{S^{-}_{i}}{S^{-}_{i}+S^{+}_{i}}, i = 1, \dots, m.$ 
A higher $C_i$ indicates that the node is closer to the positive ideal solution and farther from the negative ideal, thus signifying greater importance. 
Several studies have applied and extended this methodology. For instance,  Liu et al.~\cite{liu2015node} used degree, betweenness, closeness, and $k$-shell values to construct and standardize a decision matrix for TOPSIS-based node evaluation. Hu et al.~\cite{hu2016modified} incorporated the SIR model to adjust attribute weights and optimize the TOPSIS model. 
Other researchers have tailored TOPSIS to specific network types. Li et al.~\cite{li2023identifying23} combined resistance centrality with a weighted TOPSIS approach for weighted power grids. Dong et al.~\cite{dong2022cpr34} defined a communication probability matrix based on shortest path length and number of paths between nodes, using relative entropy to measure differences in communication distributions in their TOPSIS system. These adaptations highlight TOPSIS's flexibility and effectiveness in evaluating node importance across diverse network types.

The entropy weight method and empowerment method enhance the highest attribute in the comprehensive evaluation system, increasing the distinction of relative tightness. Chen et al.~\cite{chen2021effects} introduced an adjustable TOPSIS method to address the entropy weight method's tendency to overemphasize attributes with high data diversity. This method incorporates a weight coefficient to modulate the influence of entropy weights on the evaluation outcomes.
Ishfaq et al.~\cite{ishfaq2022identifying45} employed the entropy weight technique to assign objective weights to each criterion and then applied TOPSIS to rank nodes. Yang et al.~\cite{yang2018multi} proposed a local centrality indicator that incorporates multi-layer neighbor and clustering coefficient information. When combined with grey relational analysis (GRA) and the SIR model, this indicator more accurately identifies key nodes compared to traditional weighted TOPSIS. Vega-Oliveros et al.~\cite{vega2019multi} introduced the multi-centrality index (MCI) for document keyword extraction by optimizing the combination of centrality indicators. Lu et al.~\cite{lu2020node} used relative entropy and gray relational degree to improve node evaluation in power networks. Zhang et al.~\cite{zhang2022multi23} proposed a Multi-attribute Critic Network Decision Indicator (MCNDI) that integrates the H-index, closeness centrality, and other metrics using the objective CRITIC method for weighting. Ju et al.~\cite{ju2021novel23} introduced the Multi-Criteria Compromise Ranking Method (VIKOR) combined multiple rail transit networks into a multi-layer regional network, determined centrality criterion weights using an objective method.

\section{Information Propagation Based Ranking Methods} 
Ranking methods based on information propagation assess node importance by simulating and analyzing the spread of information or disease across the static network as illustrated in Table.~\ref{table08}. These methods offer valuable insights into dynamic spreading processes and are widely applied in social network analysis, public health disease control, and marketing strategies. Their main advantage lies in capturing both structural and dynamic aspects of diffusion, enabling a more realistic evaluation of node information maximization. However, they also face several challenges, including the difficulty of accurately modeling propagation dynamics, high computational demands for large-scale networks, and the need for domain-specific knowledge to interpret results effectively.

\subsection{Diffusion Model}\label{sec70}
Diffusion-based ranking methods assess node influence by modeling information or disease spread. Macdonald et al.~\cite{macdonald2012spreaders} applied the SIR model to compare centrality metrics, demonstrating that eigenvector centrality effectively identifies high-impact spreaders above a critical infection threshold. Borgatti et al.~\cite{borgatti2006identifying} introduced the KPP-POS and KPP-NEG algorithms: KPP-POS optimizes information flow by maximizing average diffusion efficiency, while KPP-NEG disrupts network connectivity by removing key nodes, enhancing influence identification for targeted dissemination.
Zhuge et al.~\cite{zhuge2010topological} proposed Topological Centrality (TC), iteratively updating node and edge weights to quantify influence, with a stable weight of 1 indicating centrality. Aral et al.~\cite{aral2012identifying} applied randomized experiments and a continuous-time proportional hazards model to distinguish individual influence from peer susceptibility. Addressing location-aware influence maximization, Li et al.~\cite{li2014efficient} developed approximation-ratio greedy algorithms: a boundary greedy method leveraging upper and lower bounds for Top-$k$ spreader identification, and a hint-based approach precomputing clues for efficient high-impact node detection in real-time dissemination.

To overcome the limitations of traditional centrality metrics in assessing the spreading capability of non-topologically dominant nodes, Lawyer et al.~\cite{lawyer2015understanding} introduced the Expected Force (EXP) metric, which incorporates entropy to model the distribution of infectiousness across multiple propagation rounds. Similarly, Chen et al.~\cite{chen2015critical} proposed three diffusion-based indicators—NEGD, diffusion speed, and diffusion scale—where NEGD, defined as the ratio of Expected Geodesic Distance (EGD) to Largest Geodesic Distance (LGD), quantifies overall propagation efficiency. Although these methods effectively identify influential nodes within local neighborhoods, they do not guarantee global optimality.
Robinaugh et al.~\cite{robinaugh2016identifying} improved influence estimation by introducing single-step and multi-step expected impact measures for analyzing psychopathology networks. Bozorgi et al.~\cite{bozorgi2016incim} proposed the INCIM algorithm, which combines local propagation dynamics with global network structure to evaluate both inter- and intra-community influence under the linear threshold model. Meanwhile, Holme et al.~\cite{holme2017three} investigated outbreak size, vaccination effects, and information diffusion using SIR simulations.
Collectively, these diffusion-based approaches simulate the spread of information or disease to assess node influence, integrating both local and global network characteristics to offer practical solutions for identifying critical nodes in diverse network environments.

Recent advances further enhance influence estimation. Yin et al.~\cite{yin2023identifying47} leveraged Compressed Sensing Theory (CST) to frame node identification as sparse signal reconstruction, applying the SIR model for efficient target detection. He et al.~\cite{he2019tifim} proposed the TIFIM framework, combining descending iteration and top-advantage metrics for seed node selection. Tulu et al.~\cite{tulu2019vital} developed the Node Willingness and Influence (NWI) algorithm, integrating network structure and interaction frequency for key spreader detection.
Zhong et al.~\cite{zhong2021information} introduced the Improved Information Entropy (IIE) method, leveraging higher-order neighborhoods and weighted entropy to refine node importance. Li et al.~\cite{li2021estimating} employed a Spearman-like correlation to enhance ranking precision, outperforming degree and betweenness centrality. Gong et al.~\cite{gong2021probability} proposed the Probability-Driven Structure-Aware (PDSA) algorithm, which updates influence estimates via graph traversal in the IC model. Wang et al.~\cite{wang2023hgim} introduced the HGIM method, constructing a heterogeneous propagation graph to integrate topology with diffusion cascades. To model dynamic influence, Mohammadi et al.~\cite{mohammadi2024improved} proposed the Two-Sided Sign-Aware Matching (TSM) framework, incorporating trust and reciprocity in signed networks. Wang et al.~\cite{wang2024identifying} introduced the Dynamic Propagation Probability (DPP) model, redefining neighbor influence within a three-hop neighborhood. Xu et al.~\cite{xu2023novel} developed the Local Propagation Probability (LPP) model, integrating hierarchical propagation influence across different orders.

For large-scale networks, Ai et al.~\cite{ai2022identifying10} proposed the SPC method, which balances network connectivity with propagation precision by defining infection probabilities within a truncated radius. Zareie et al.~\cite{zareie2020identification} optimized key node selection using the Gray-golf algorithm, incorporating second-order neighbor effects for propagation efficiency. Chen et al.~\cite{chen2020influential} introduced the Dynamic Influence Seed Selection (DYISSE) method, integrating two-hop triangular influence measures for robust diffusion estimation.
Fink et al.~\cite{fink2023centrality} introduced virus centrality, leveraging weighted directed networks to estimate propagation potential. Sun et al.~\cite{sun2023finding} further refined IDME by integrating neighborhood similarity and centrality. Ullah et al.~\cite{ullah2023lss} proposed the Local Structure System (LSS), which incorporates $k$-shell, degree, and triangle count to quantify influence. Finally, Corsin et al.~\cite{corsin2025evidence} developed an agent-based information diffusion model, demonstrating that source persistence has a greater impact on network consensus than source quantity, particularly in polarized environments.

\subsection{Dynamical Sensitivity}\label{sec71}
Liu et al.~\cite{liu2016locating} has proposed the dynamic sensitive centrality (DSC) by integrating topological features and dynamic characteristics, considering a discrete time spreading model in which an infected node will infect its neighbors with a spreading rate $\beta$ and recover with a recovery rate $\mu$. Let $x(t)(t>0)$ denote the 
approximate value of the cumulative probability of nodes being activated between time step 1 and $t$, then $x(t)-x(t-1)(t>1)$ is approximated as the probability of a node being 
infected at time step $t$. If $v_i$ is the unique initial infected node, then $x_{i}(0)=1$ and $x_{j\ne i}(0)=0$. In the first time step, $x(1)=\beta Ax(0)$, and for $t>1$: $ x(t)-x(t-1) = \beta A[\beta A+(1-\mu)I]^{t-1}x(0)$, where $v_i$ is the identity matrix. Let $H=\beta A +(1-\mu)I$, then the cumulative probability of nodes being infected between time step 1 and $t$ can be approximated as
\begin{equation}
 x(t) = \sum_{r=2}^{t}[x(r)-x(r-1)]+x(1) = \sum_{r=0}^{t-1}~\beta A H^{r}x(0).
\end{equation}
Define $S_{i}(t)$ as the spreading influence of node $i$ at time step $t$. In the case where node $i$ is initially infected, it can be quantified as the sum of the infection probabilities of all nodes. The infection probability in the equation above is $x(t) = \sum_{r=0}^{t-1}~\beta A H^{r} e_{i}$ where $e_{i}=(0, \dots, 0, 1, \dots, 0)^{T}$ is a vector with only the $i$-th entry being 1. Therefore, the spreading influence of node $v_i$ can be defined as $S_{i}(t) = [(\beta A +\beta AH + \dots + \beta AH^{t-1})^{T}L]$ where $L=(1, 1, \dots, 1)^{T}$ is a vector of size $n\times 1$. DS centrality directly measures node influence, capturing both structural and dynamic factors.
Mo et al.~\cite{mo2019identifying} integrated multiple centrality metrics—including degree, betweenness, efficiency, and correlation—using the Dempster-Shafer (D-S) evidence theory. By treating these metrics as distinct probability assignments and applying evidence combination rules, they developed a comprehensive indicator for node importance.
Du et al.~\cite{du2020identifying} introduced an Improved Topological Potential Model (ITPE) that leverages topological entropy to assign weights to centrality measures such as weighted degree, betweenness, closeness, and eigenvector centrality. By optimizing control parameters to minimize topological entropy.

\begin{table}[htbp]
 \centering
 \caption{Comparison of Comprehensive Index-Based and Information Propagation Based Ranking Methods.}
\label{table08}
 \begin{tabular}{m{4cm}m{4cm}m{7cm}}
 \toprule
 Title & Related Works & Advantages / Disadvantages\\
 \midrule Gravity Formula
 & \cite{RN14},
~\cite{maji2020influential},
~\cite{RN16},
~\cite{RN24},
~\cite{liu2020gmm},
~\cite{RN13},
~\cite{RN20},
~\cite{fei2018identifying},
~\cite{RN28},
~\cite{RN26},
~\cite{yan2020identifying},
~\cite{wang2022identifying20},
~\cite{yang2023aigcrank17},
~\cite{zhu2023identifying12},
~\cite{liu2023entropy12},
~\cite{zhao2022random21},
~\cite{zhao2023estimating3},
~\cite{RN25},
~\cite{curado2023novel},
~\cite{yang2023aogc},
~\cite{chen2023identification},
~\cite{meng2025improved},
~\cite{xu2024cagm},
 ~\cite{li2023improved123},
~\cite{lv2024improved},
~\cite{chi2024measuring}

 & 
 $+$ Comprehensive consideration of multiple factors, flexibility in handling various network structures.
 
 $-$ Potential complexity in parameter tuning, reliance on accurate index weightings.
 \\
 \midrule Topsis and Entropy-Weight
 
 Multiple Metrics
 &\cite{comin2011identifying},
~\cite{de2014role},
~\cite{hu2015multi},
~\cite{hu2015identifying},
~\cite{bucur2020top},
~\cite{wei2022identifying24},
~\cite{an2024novel},
~\cite{wu2024hunting},
~\cite{cao2024dynamic},
~\cite{kopsidas2022identification},
~\cite{wang2024multi},
~\cite{lei2024weighted},
~\cite{ullah2024leveraging},
~\cite{zhang2024critical},
~\cite{lee2024identifying},
~\cite{esfandiari2025collaborative},
~\cite{chen2025sfimco},
~\cite{zhang2025novel},
~\cite{mo2015evidential},
~\cite{xu2025lgp},
~\cite{sheikhahmadi2017identification12},
~\cite{wang2017ranking},
~\cite{sheikhahmadi2017identification},
~\cite{yang2018ranking},
~\cite{ullah2021identification},
~\cite{ullah2021identifying12},
~\cite{hu2022exploring209},
~\cite{wang2022influential},
~\cite{mukhtar2023integrating},
~\cite{yang2023identifying23},
~\cite{qiu2024key},
~\cite{du2014new},
~\cite{liu2015node},
~\cite{hu2016modified},
~\cite{li2023identifying23},
~\cite{dong2022cpr34},
~\cite{chen2021effects},
~\cite{ishfaq2022identifying45},
~\cite{yang2018multi},
~\cite{vega2019multi},
~\cite{lu2020node},
~\cite{zhang2022multi23},
~\cite{ju2021novel23}
 & $+$ Holistic analysis by integrating multiple evaluation criteria, increases accuracy in identifying critical nodes, adaptability to different network types. 
 
 $-$ Difficulties in balancing and weighting diverse criteria effectively.

 \\
 \midrule Information propagation method
 &\cite{macdonald2012spreaders},
~\cite{borgatti2006identifying},
~\cite{zhuge2010topological},
~\cite{aral2012identifying},
~\cite{li2014efficient},
~\cite{lawyer2015understanding},
~\cite{chen2015critical},
~\cite{robinaugh2016identifying},
~\cite{bozorgi2016incim},
~\cite{holme2017three},
~\cite{yin2023identifying47},
~\cite{he2019tifim},
~\cite{tulu2019vital},
~\cite{zhong2021information},
~\cite{li2021estimating},
~\cite{gong2021probability},
~\cite{wang2023hgim},
~\cite{mohammadi2024improved},
~\cite{wang2024identifying},
~\cite{xu2023novel},
~\cite{ai2022identifying10},
~\cite{zareie2020identification},
~\cite{chen2020influential},
~\cite{fink2023centrality},
~\cite{sun2023finding},
~\cite{ullah2023lss},
~\cite{corsin2025evidence},
~\cite{liu2016locating},
~\cite{mo2019identifying},
~\cite{du2020identifying}
 & $+$ Effective identification of nodes crucial for information flow, applicability to diverse network types. 
 
 $-$ Potential computational intensity, reliance on accurate modeling of propagation processes.
 \\
 \bottomrule
 \end{tabular}
\end{table}

\section{High-order Networks Based Ranking Methods}
\begin{figure}[h]
    \centering
    \includegraphics[width=1\linewidth]{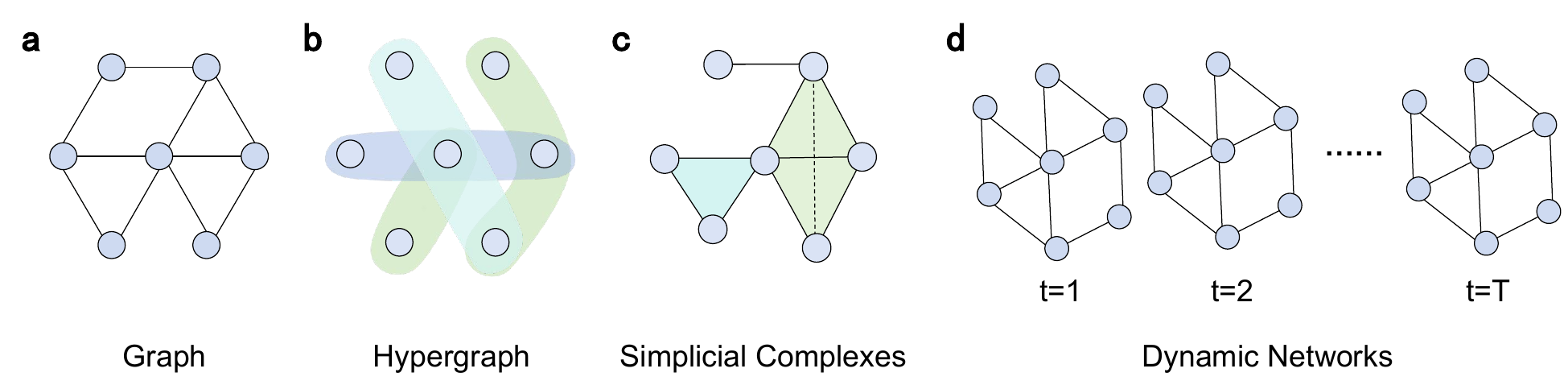}
    \caption{Network Structure Illustration.}
    \label{fig02}
\end{figure}

Unlike traditional pairwise connections as Graph in Figure.~\ref{fig02}~(a) , high-order networks, such as simplicial complexes and hypergraphs as shown in Figure.~\ref{fig02}~(b)(c). Centrality measures typically used in conventional networks can be extended by integrating higher-order relationships, leading to the development of novel node importance metrics. Recent advancements in hypergraph-based ranking methods have improved node importance estimation by integrating higher-order relationships.
Kapoor et al.~\cite{kapoor2013weighted} extended degree centrality to hypergraphs, defining weighted degree centrality as the sum of weights of hyperedges connecting a node to others. Specifically, Given a hypergraph $H=(V, E)$, the weighted degree centrality of a node is defined as
\begin{equation}
C^{h}_{d}(i)=\sum_{j=1, j\ne i}^{N}~\sum_{k=1, \{v_{i}, v_{j}~\} \subset e_{k} }^{L} w_{k}, 
\end{equation}
where $w_{k}$ is the weight of the hyperedge $e_{k}$. Furthermore, Kapoor et al.~\cite{kapoor2013weighted} introduced a distinction between strong and weak ties based on Granovetter's concept of tie strength, which enables more accurate node importance estimation.
In a similar vein, Lee et al.~\cite{lee2021betweenness} introduced hyperedge betweenness centrality, adapting traditional betweenness measures for hypergraphs by converting them into bipartite graphs. This method refines the understanding of node influence in complex networks.
Building on these methods, St-Onge et al.~\cite{st2022influential} developed an approximate master equation to study influence propagation in heterogeneous hypergraphs. They quantified node spreading potential using hyperedge cardinality and hyperdegree, identifying key seed nodes.
Xie et al.~\cite{XIE2023103161} proposed HADP, which prioritizes nodes with the highest initial fitness as seed set members. It constructs and updates an adaptive set of seed node neighbors, applying degree penalty terms across all nodes in large-scale networks. This approach applies degree penalty terms across all nodes in large-scale networks, thus avoiding the complexity of individually calculating the fitness of each node. Lee et al.~\cite{Lee2023mmlmathXA} and Mancastroppa et al.~\cite{Mancastroppa2023HypercoresPL} proposed decomposing hypergraphs into hypercores, where nodes belong to at least k hyperedges of size at least m. This method identifies localized diffusing nodes more effectively than core degree metrics.

In terms of hypergraph eigenvector centrality, Benson et al.~\cite{Benson_Austin_R} extended eigenvector centrality to uniform hypergraphs using tensor Perron-Frobenius theory. Zhao et al.~\cite{zhao2023ranking} introduced higher-order centrality measures incorporating low-order network metrics. These metrics incorporate low-order network metrics to more accurately assess the relative importance of clusters in higher-order structures.
Li et al.~\cite{li2023influence} proposed an electrostatic field-based algorithm to identify seed node sets in hypergraphs. Xie et al.~\cite{xie2023vital} introduced LHGC, which incorporates node degree and higher-order distance in a gravitational model.
The lack of research on higher-order relationships in dynamic networks has been addressed by Zhao et al.~\cite{zhao2023general} developed centrality matrices for Pagerank, Hub, and Authority by adjusting weights of network motifs at different orders.

Beyond binary network models, simplicial complexes, composed of $k+1$ vertices forming $k$-dimensional simplexes such as edges, triangles, and tetrahedra, offer a natural framework to capture higher-order interactions in complex systems~\cite{battiston2020networks}. By incorporating simplexes of different orders, these structures allow a more detailed representation of the intricate relationships within networks, as highlighted by Goldberg~\cite{goldberg2002combinatorial} and Estrada et al.~\cite{estrada2018centralities}.
Building on this framework, Serrano et al.~\cite{serrano2020simplicial} introduced generalized adjacency measures based on the intersection patterns among simplices, extending traditional centrality concepts to higher-order contexts. Their approach, which includes random walk-based measures over sequences of $k$-order simplexes, enables the characterization of structural features such as shortest path lengths, eccentricity, and network diameter~\cite{granovetter1973strength}. This methodology offers a comprehensive perspective for probing both global structure and local properties.
Furthermore, several studies have adapted classical graph-theoretic methods to simplicial complexes. Notably, adjacency matrices defined over $k$-order simplexes have been employed to compute eigenvector centrality~\cite{bonacich2007some} and Katz centrality~\cite{estrada2015first}, thus providing tools to capture both local connectivity patterns and global topological features. Overall, simplicial complex models have emerged as a versatile and powerful framework for the analysis of complex networks, enabling deeper insights into their underlying interactions and dynamic behaviors.

\section{Dynamic Networks Based Ranking methods}
Recent research has focused on developing ranking methods for dynamic networks (Figure.~\ref{fig02}(d)) to better capture the evolving nature of node interactions and improve node importance estimation. Kim et al.~\cite{kim2012temporal} introduced the time-ordered graph to simplify dynamic networks into static networks of directed flow. They expanded traditional centrality measures, such as degree, betweenness, and closeness to dynamic graphs using the following definitions:
\begin{equation}
\begin{aligned}
 D_{i, j}(v) &=\sum_{t=i}^{j} 2D_{t}(v), \quad
 C_{i, j}(v) =\sum_{i\le t<j}~\sum_{u\in V\setminus v} \frac{1}{\Delta_{t, j}(v, u)}, \\
 B_{i, j}(v) &=\sum_{i\le t <j}~\sum_{s\ne v \ne d \in V, \sigma_{i, j}(s, d)>0} \frac{\sigma_{t, j}(s, d, v)}{\sigma_{t, j}(s, d)}, \\
\end{aligned}
\end{equation}
where $D_i(v)$ represents node $v$ at time $i$, $\Delta_{t, j}(v, u)$ is the shortest path length between nodes $v, u$ in the time interval $[t, j]$, $\sigma_{i, j}(s, d)$ refers to the shortest path length between nodes $s, d$ in the time interval $[i, j]$, and $\sigma_{t, j}(s, d, v)$ refers to the shortest path length containing $u$ in the path between nodes $s, d$ in the time interval $[i, j]$.
Tsalouchidou et al.~\cite{tsalouchidou2020temporal} highlighted the significance of temporal betweenness centrality (TBC) in influencing information spread in dynamic networks. They proposed a bi-objective algorithm for the shortest-fastest path (SFP) in temporal graphs, combining spatial and temporal factors to define TBC. This metric is sensitive to spatial and temporal distances between vertices, enhancing the understanding of communication intermediaries in temporal networks.
Similarly, Elmezain et al.~\cite{elmezain2021temporal} applied the Opshal algorithm to calculate degree centrality and shortest distances in weighted networks, defining metrics such as temporal degree and temporal closeness to assess node importance over time.

In the realm of multilayer temporal networks, Lü et al.~\cite{lv2019eigenvector} developed the ECMSim method, utilizing a fourth-order tensor to capture node similarity and PageRank for tensor-based networks. They further refined this approach with MTEIGBC and MTPRBC centralities~\cite{lv2019pagerank}, which employ cosine similarity to rank nodes across multiple dimensions including layers and timestamps.
Expanding upon these concepts, Lü et al.~\cite{lv2021hits} proposed the MT-HITS centrality. This method incorporates inter-layer similarity coefficients and solves tensor equations to define centrality vectors across multiple dimensions.
Bi et al.~\cite{bi2021temporal} adapted the gravitational model to dynamic networks through the Temporal Gravity Model (TGM). TGM defines temporal distance based on both network structure and temporal order, integrating static centrality measures with their dynamic extensions. 

To address the limitations of fixed constants in multilayer coupled networks, Jiang et al.~\cite{jiang2022identifying} developed the enhanced similarity index (ESI) to assess interlayer coupling in multilayer networks, moving beyond fixed constants. By integrating node neighbors across two time layers and introducing an attenuation factor, they created a decay-based super adjacency matrix (ASAM) to identify key node pairs affecting the largest connected component in temporal networks.
Taylor et al.~\cite{taylor2017eigenvector} integrated centrality matrices from different network layers into a joint $NT \times NT$ matrix, capturing node and time layer importance. They defined marginal and conditional centrality, leading to a time-averaged eigenvector centrality that emphasizes strong interlayer connections.
Tao et al.~\cite{tao2022sequential} introduced sequential path trees, which require at least one time-related path between the root and any subnode. These trees incorporate propagation time, intermediate nodes, and reachable paths. A normalized feature matrix based on centrality in these trees quantifies node spreading ability. Chen et al.~\cite{chen2024temporal} extended the local clustering coefficient to dynamic networks, revealing hidden connectivity through the temporal local clustering coefficient.
Pan et al.~\cite{pan2011path} highlighted the role of temporal distance in information dissemination speed in dynamic networks. Their findings highlight that dissemination speed is constrained by the average temporal distance between nodes, emphasizing the critical role of the temporal dimension in understanding network dynamics.

To track influential nodes in dynamic social networks, Song et al.~\cite{song2016influential} framed the problem as influence maximization in dynamic environments. They proposed an upper-limit exchange greedy algorithm, which optimizes influence coverage by replacing nodes in the previously identified set of influential nodes, thus enabling efficient continuous tracking as the network evolves.
Building on these ideas, Huang et al.~\cite{huang2017dynamic} extended dynamic sensitivity to temporal networks by introducing Temporal Dynamic Sensitivity Centrality (TDC), which highlights that both the topology and dynamics of the network affect node spread. Additionally, Huang et al.~\cite{huang2017centrality} defined a super-evolution matrix to model temporal network structures, effectively reducing computational complexity by transforming the problem of calculating eigenvector centrality into the computation of eigenvectors from low-dimensional matrices. Additionally, Qu et al.~\cite{qu2019temporal} introduced the Temporal Information Gathering (TIG) process, which defines node importance based on four key factors: time information gathering depth, time distance matrix, initial information, and a weighting function. Furthermore, Arebi et al.~\cite{arebi2022effective} proposed a method based on temporal centrality measures to identify the minimum driver nodes set (MDS) that can fully control the network. Meanwhile, Li et al.~\cite{Li10317077} proposed a temporal information fusion model, which successfully identifies critical nodes in dynamic power information networks from both a topological and informational perspective.

To address the high settlement cost of temporal Katz centrality, Zhang et al.~\cite{zhang2024tatkc} introduced the Two-stage Attention-based Temporal Knowledge Graph Convolutional Network (TATKC). This framework combines temporal attention mechanisms with MLP to predict node ranking, enhancing performance through normalization and neighbor node sampling strategies. 
Similarly, Yu et al.~\cite{yu2020identifying22} constructed a feature matrix based on neighborhood information and used the SIR model to identify key nodes in temporal networks. By employing network embedding and machine learning, they proposed the MLI algorithm, which transforms the identification of key nodes into a regression problem. In addition, Yu et al.~\cite{yu2023predicting} proposed a framework that combines GCN and RNN to identify nodes with optimal propagation capabilities, considering the sequential topology of temporal networks. Furthermore, Wang et al.~\cite{wang2024influential} modeled node propagation as a hierarchical infection process, proposing Hierarchical Structure Influence (HSI), which quantifies a node's propagation potential across the entire network.

\section{Summary and Discussion}

In this review, we have systematically examined the concepts and methodologies associated with identifying important, critical, influential, and key nodes in complex networks and systems. Existing research has been broadly categorized into three main classes: structurally important nodes, highly influential nodes in information spreading, and critical nodes under performance constraints.
In static networks, traditional centrality measures, such as degree, betweenness, and clustering coefficient, have provided the foundation for assessing node importance. However, the advent of dynamic, multilayer, and higher-order network models has given rise to a new generation of metrics, including spatio-temporal coupling indicators, propagation-based dynamic centralities, and hypergraph-based approaches. These methods have deepened our understanding of node roles and spreading mechanisms, particularly in contexts where interactions evolve over time and across multiple layers. Furthermore, models that incorporate coupling effects, edge weight dynamics, temporal features, and information fusion have enabled more robust identification of critical nodes in complex, real-world settings. Such advances have demonstrated practical relevance across diverse domains, including social networks, biological networks, power grid networks, transportation networks, and geographical networks.

Despite these developments, significant challenges persist in accurately pinpointing key nodes within real-world networks characterized by heterogeneity and dynamic complexity. The following areas require fulture direction: 
(1)~\textbf{Dynamic Network Evolution Patterns:} Existing methods often merely extend static indicators to dynamic networks, with insufficient study on the evolution patterns of key nodes and their spatial distribution in dynamic networks. This limits the exploration of potential patterns in complex systems. Additionally, researchers should focus more on the identifying and inferring higher-order structures and dynamics in higher-order networks.
(2)~\textbf{Efficient Algorithm Development:} As networks grow in scale and complexity, there is an urgent need for efficient algorithms with low computational overhead. Many existing heuristic methods for node removal lack scalability and are ill-suited to large, dynamically evolving systems. Research should focus on developing real-time tracking algorithms, influence estimation techniques, and control strategies tailored to dynamic and multilayer networks.
(3)~\textbf{Integration of Machine Learning and Data-Driven Approaches:} Recent advances in machine learning and data-driven methods, such as network embedding and deep learning, provide new perspectives for node importance identification. The emergence of novel network data structures like semantic text attribute graphs and knowledge graphs offers opportunities to combine traditional centrality measures with modern machine learning models. This integration can enhance prediction accuracy and adaptability in recommendation systems, databases, and other applications.
(4)~\textbf{Cross-Domain Applications and Validation:} Most current node importance measures are context-specific, limiting their generalizability. Developing a unified framework that accounts for structural, dynamic, and higher-order features is essential. Different domains (e.g., social networks, biological systems, power grids, transportation networks) have distinct definitions and requirements for key nodes. Future research should emphasize interdisciplinary collaboration, validate theoretical models with real-world datasets, and develop customized solutions for specific scenarios.
In conclusion, the identification of key nodes in complex networks is a vibrant research area with significant theoretical and practical implications. Addressing the aforementioned challenges and future directions will contribute to a deeper understanding of complex systems and enable more effective solutions for real-world problems.

\newpage
\begin{landscape}
\begin{table}[htbp]
\small
\caption{Overview of article structure, method categorization, and main representative approaches.}
\label{table02}
\renewcommand{\arraystretch}{1.1} 
\begin{tabular}{m{4cm}m{7cm}m{11cm}}
\hline 
\hline 
Ranking Methods & Categories & Representative Methods \\
\hline 
\multirow{6}{*}{Neighbors-Based } & \nameref{sec10} & DC~\cite{RN3}, SLC~\cite{RN4},LC~\cite{ma2019quasi}, ILC~\cite{zhu2024identifying} \\
 & \nameref{sec11} & K-shell~\cite{RN41}, MDD~\cite{RN42},M-Centrality~\cite{RN6} \\
 & \nameref{sec12} & H-index~\cite{lu2016h}, LH-index~\cite{RN10}, EHC~\cite{zareie2019ehc} \\
 & \nameref{sec13} & K-truss~\cite{2008Trusses}, Trust-core~\cite{malliaros2016locating} \\
 & \nameref{sec14} &SH~\cite{yu2017critical}, ISH~\cite{yu2017identifying},
VKC~\cite{xu2018identifying},
ICC~\cite{liu2021identifying54}, SHKS~\cite{zhao2023ranking21} \\
 & \nameref{sec15} & HP~\cite{wang2015link},
 ESCRM~\cite{lu2022critical11},
DNS~\cite{rao2022cbim},
TC~\cite{tong2024novel} \\
\hline 
\multirow{3}{*}{Path} & \nameref{sec20} & BC~\cite{RN56}, ECC~\cite{hage1995eccentricity}, ASP~\cite{RN59}, RASP~\cite{RN61},LC~\cite{song2015novel},CNICC~\cite{ventresca2015efficiently}
EOCI~\cite{yang2020critical}, , LASPN~\cite{xiao2024new} \\
 & \nameref{sec21} & CC~\cite{freeman2002centrality},EFF~\cite{latora2001efficient},GLR~\cite{salavati2019ranking},GLS~\cite{sheng2020identifying} \\
 & \nameref{sec22} & RW~\cite{RN7}, DRW~\cite{iannelli2018influencers} 
 \\
\hline 
\multirow{4}{*}{Eigenvector} & 
\nameref{sec30}
&EC~\cite{bonacich1971factoring},PCC~\cite{ilyas2011identifying}, LeadersRanK~\cite{ahajjam2015leadersrank},ECDS~\cite{zhong2018identifying}
 \\
 & \nameref{sec31} & Katz~\cite{katz1953new} \\
 & \nameref{sec32} & Pagerank~\cite{page1998pagerank},AttriRank~\cite{hsu2017unsupervised}, TPR~\cite{sheng2020identifying12}, EPR~\cite{su2021identification67} \\
 & \nameref{sec33} & LeaderRank~\cite{RN33}, WLeaderRank~\cite{li2014identifying},Accumulated Nomination~\cite{poulin2000dynamical},VoteRank~\cite{RN36} \\
\hline 
\multirow{3}{*}{Control-Optimization } 
& \nameref{sec43} & R-PGME~\cite{ding2017key}, BCNs~\cite{lu2019pinning}, ~\cite{li2006controlling},
~\cite{yu2012distributed},
~\cite{yu2013synchronization},
~\cite{sun2025optimal},
~\cite{zhang2017effect},
~\cite{mo2019effects},
~\cite{jiang2023searching},
~\cite{cao2024synchronization},
~\cite{zhang2022necessary},
 ~\cite{gao2025effects}   \\
& \nameref{sec40} & CC-CNDP~\cite{Boginski0007}, CC-CNP~\cite{arulselvan2011cardinality},BPT~\cite{li2011finding},MOCNDP~\cite{ventresca2015experimental},DSFLA~\cite{tang2020discrete} 
 \\ 
 &\nameref{sec41}& ILP~\cite{di2012branch} ,EPC~\cite{dinh2015assessing},GRASP~\cite{purevsuren2016heuristic}, HHA~\cite{addis2016hybrid} ,MIQP~\cite{chen2020critical},HILPR~\cite{shen2012discovery} ,RGDC~\cite{ventresca2014region}, MSD~\cite{yin2023mixed23}, BICND~\cite{zhang2024evolutionary}, 
EBC~\cite{jiang2025identifying} \\
 & \nameref{sec42} &MST~\cite{chen2004evaluation} ,CNC~\cite{hermelin2016parameterized}, CPND~\cite{addis2013identifying} ,MCDS~\cite{aringhieri2016local} \\

\hline 
\multirow{4}{*}{Machine Learning} & \nameref{sec50} & DSR~\cite{zareie2019influential},
ME~\cite{nitt2016using},
EnRenew~\cite{guo2020influential},
ODEF~\cite{xu2018discovery},
CbM~\cite{tulu2018identifying},
EMI~\cite{li2019key} \\
 & \nameref{sec51} & ClusterRank~\cite{RN35},
LSC~\cite{gao2014ranking},
ECRM~\cite{yang2018identifying},
ACE~\cite{dablander2019node},
GDF-ICN~\cite{liu2022identifying11}~\\
 & \nameref{sec52} & NLC~\cite{yang2021identifying36},
SVR~\cite{rezaei2023machine356},
InfGCN~\cite{ZHAO202018},
CGNN~\cite{zhang2022new57},
DeepIM~\cite{keikha2020influence},
ETIM~\cite{bouyer2024maximizing},
LCNN~\cite{ahmad2024learning},
AGNN~\cite{xiong2024vital},
RCNN~\cite{yu2020identifying},
NIE-GAT~\cite{liu2022nie23},
MFA-NIE~\cite{ge2024node},
GENI~\cite{park2019estimating},
MultiImport~\cite{park2020multiimport},
ILGR~\cite{munikoti2022scalable},
MCRL~\cite{liu2023node},
AGCL~\cite{shu2025node},
MCRL~\cite{liu2023node},
LICAP~\cite{zhang2024label},
HIVEN~\cite{huang2022estimating},
SKES~\cite{chen2024deep},
LENIE~\cite{lin2024node},
HONNMA~\cite{zhao2023novel}
\\
 & \nameref{sec53} & FINDER~\cite{fan2020finding},
DeepELE~\cite{jaques2019social},
DISCO~\cite{li2019disco},
ToupleGDD~\cite{chen2023touplegdd},
DeepIM~\cite{ling2023deep},
PIANO~\cite{li2022piano},
DERL~\cite{uthayasuriyan2024adaptive},
AdaRisk~\cite{li2024adarisk},
HEDRL-IM~\cite{xu2025influence},
BiGDN~\cite{zhu2025bigdn},
CoreQ~\cite{ahmad2025learning} \\
\hline 
\multirow{4}{*}{Comprehensive Index-Based } & \nameref{sec60} & GM~\cite{RN14},
LGM~\cite{RN16}GMM~\cite{liu2020gmm},
IGM~\cite{RN28},
MCCM~\cite{RN26},
AIGCrank~\cite{yang2023aigcrank17},
CAGM~\cite{xu2024cagm},
HVGC~\cite{zhu2023identifying12},
SEGM~\cite{liu2023entropy12},
NEGM~\cite{zhao2023estimating3},
DCGM~\cite{chen2023identification},
ACM~\cite{meng2025improved},
APAMGM~\cite{li2023improved123} \\
 & \nameref{sec61} & PRGC~\cite{lv2024improved},
IMC~\cite{tan2006evaluation},
KSC~\cite{HuQingcheng2013},
RWA~\cite{de2014role},
NICCM~\cite{hu2015identifying},
DIRCI~\cite{an2024novel},
INLGC~\cite{cao2024dynamic},
CNT~\cite{kopsidas2022identification},
WII~\cite{lei2024weighted},
NPIC~\cite{ullah2024leveraging},
CVR~\cite{zhang2024critical},
BII~\cite{lee2024identifying},
HNPR~\cite{esfandiari2025collaborative},
SFIMCO~\cite{chen2025sfimco},
SLCMF~\cite{zhang2025novel},
CEC~\cite{mo2015evidential},
LGP-DS~\cite{xu2025lgp},
MCDE~\cite{sheikhahmadi2017identification12},
NGSC~\cite{namtirtha2021best} \\
 & \nameref{sec62} & TOPSIS~\cite{du2014new},
Entropy-TOPSIS~\cite{dong2022cpr34},
GRA~\cite{yang2018multi},
MCI~\cite{vega2019multi},
MCNDI~\cite{zhang2022multi23},
VIKOR~\cite{ju2021novel23}~\\
\hline 
\multirow{2}{*}{Information Propagation-Based } & \nameref{sec70} & KPP-POS~\cite{borgatti2006identifying},
TC~\cite{zhuge2010topological},
EXP~\cite{lawyer2015understanding},
NEGD~\cite{chen2015critical},
INCIM~\cite{bozorgi2016incim},
CST~\cite{yin2023identifying47},
 TIFIM~\cite{he2019tifim},
NWI~\cite{tulu2019vital},
IIE~\cite{zhong2021information},
PDSA~\cite{gong2021probability} ,
HGIM~\cite{wang2023hgim},
TSM~\cite{mohammadi2024improved},
DPP~\cite{wang2024identifying},
LPP~\cite{xu2023novel},
SPC~\cite{ai2022identifying10},
DYISSE~\cite{chen2020influential} ,
IDME~\cite{sun2023finding},
LSS~\cite{ullah2023lss}~\\
 & \nameref{sec71} & DS~\cite{liu2016locating}, 
ITPE~\cite{du2020identifying} \\
\hline 
\multicolumn{2}{l}{High-order Networks} & DC~\cite{kapoor2013weighted} , 
BC~\cite{lee2021betweenness} ,
HADP~\cite{XIE2023103161},
HyperCore\cite{Mancastroppa2023HypercoresPL} ,
CEC~\cite{Benson_Austin_R},
HOC~\cite{zhao2023ranking},
EFS~\cite{li2023influence},
LHGC~\cite{xie2023vital} , 
Pagerank~\cite{zhao2023general}
 \\
\hline 
\multicolumn{2}{l}{Dynamic Networks} & TDC~\cite{kim2012temporal}, 
TBC~\cite{tsalouchidou2020temporal},
ECMSim~\cite{lv2019eigenvector, lv2019pagerank} ,
MT-HITS~\cite{lv2021hits}, 
TGM~\cite{bi2021temporal} ,
ESI~\cite{jiang2022identifying},
D-Test\cite{song2016influential},
TDC~\cite{huang2017dynamic},
TIG~\cite{qu2019temporal}, 
MDS\cite{arebi2022effective},
TATKC~\cite{zhang2024tatkc}, 
MLI\cite{yu2020identifying22},
OSAM\cite{yu2023predicting},
HSI\cite{wang2024influential} \\
\hline 
\hline 
\end{tabular}
\end{table}
\end{landscape}

\newpage
\bibliographystyle{unsrt}
\bibliography{references}

\end{document}